\documentclass[12pt,a4paper]{article}
\usepackage{jcappub}
\usepackage{newtxtext,newtxmath}
\usepackage{color,xcolor}
\usepackage{graphicx}	% Including figure files
\usepackage{latexsym,amsmath,amssymb}
\usepackage{multicol}
\usepackage{subfigure,float}
\usepackage{dblfloatfix}
\usepackage{soul}

\usepackage[normalem]{ulem}

\usepackage{silence}

\usepackage{multirow}

\DeclareRobustCommand{\VAN}[3]{#2}
\let\VANthebibliography\thebibliography
\def\thebibliography{\DeclareRobustCommand{\VAN}[3]{##3}\VANthebibliography}

\usepackage{array}
\newcolumntype{P}[1]{>{\centering\arraybackslash}p{#1}}

\DeclareUnicodeCharacter{2009}{~}
\DeclareUnicodeCharacter{2248}{~} 
\DeclareUnicodeCharacter{2243}{~}

\nocite{*}

\def \HI{{\sc Hi}}

\def \mp{{\,\rm Mpc}^{-1}}

\newcommand{\be}{\begin{equation}}
\newcommand{\e}{\end{equation}}
\newcommand{\bear}{\begin{eqnarray}}
\newcommand{\ear}{\end{eqnarray}}

\newcommand{\comment}[1]{}

\newcommand{\stkout}[1]{\ifmmode\text{\sout{\ensuremath{#1}}}\else\sout{#1}\fi}

\newcommand{\mpci}{\;{\rm Mpc}^{-1}}
\usepackage{cancel}

\title{The re-markable 21-cm power spectrum I: Probing the \HI\ distribution in the post-reionization era using marked statistics}

\author[a]{Mohd Kamran,}\emailAdd{kamranmohd080@gmail.com}
\author[a]{Martin Sahlén,}
\author[b,c,d]{Debanjan Sarkar,}
\author[e,f]{Suman Majumdar}

\affiliation[a]{Theoretical Astrophysics, Department of Physics and Astronomy, Uppsala University, Box 516, 751 20 Uppsala, Sweden}
\affiliation[b]{Department of Physics and Trottier Space Institute, McGill University, QC H3A 2T8, Canada}
\affiliation[c]{Ciela—Montreal Institute for Astrophysical Data Analysis and Machine Learning, QC H2V 0B3, Canada}
\affiliation[d]{Department of Physics, Ben-Gurion University of the Negev, Be’er Sheva 84105, Israel}
\affiliation[e]{Department of Astronomy, Astrophysics \& Space Engineering, Indian Institute of Technology Indore, Indore 453552, India}
\affiliation[f]{Department of Physics, Blackett Laboratory, Imperial College, London SW7 2AZ, U. K.}

\date{\today}

\abstract{
The neutral hydrogen (\HI) power spectrum, measured from intensity fluctuations in the 21-cm background, offers insights into the large-scale structures (LSS) of our Universe in the post-reionization era (redshift $z<6$). A significant amount of \HI\ is expected to reside in low- and intermediate-density environments, but the power spectrum mainly captures information from high-density regions. To more fully extract the information contained in the \HI\ field, we investigate the use of a marked power spectrum statistic. Here, the power spectrum is effectively re-weighted using a non-linear mark function which depends on the smoothed local density, such that low- or high-density regions are up- or down-weighted. This approach may also capture information on some higher-order statistical moments of the field. We model the \HI\ distribution using semi-numerical simulations and for the first time study the marked \HI\ power spectrum, across $1 \leq z \leq 5$. 

Our analysis indicates that there is considerable evolution of the \HI\ field during the post-reionization era. Over a wide range of length scales (comoving wave numbers $0.05\leq k \leq 1.0$ Mpc$^{-1}$) we expectedly find that the \HI\ evolves slowly at early times, but more rapidly at late times. This evolution is not well-captured by the power spectrum of the standard (unmarked) \HI\ field. We also study how the evolution of the \HI\ field depends on the chosen smoothing scale for the mark, and how this affects the marked power spectrum. 

We conclude that the information about the \HI\ content at low and intermediate densities is important for a correct and consistent analysis of \HI\ content and evolution based on the 21-cm background. The marked power spectrum can thus provide a less biased statistic for parameter constraints than the normal power spectrum.
}


\begin{document}
\maketitle
%\label{firstpage}
\vspace{2em}

%%%%%%%%%%%%%%%%%%%%%%%%%%%%%%%%%%%%%%%%%%%

\section{Introduction}

The 21-cm radiation arising from the hyperfine transition within the ground state of the neutral hydrogen atom (\HI) provides a unique means of mapping large-scale structures (LSS) over a wide range of redshifts in the post-reionization era ($z < 6$) \citep{Madau:1996cs, Furlanetto:2006jb}. The collective 21-cm emission from individual \HI\ sources in the post-reionization era creates a diffuse background radiation below $1420$ MHz. Detecting intensity fluctuations in this 21-cm background can offer insights into the underlying LSS, a technique known as 21-cm intensity mapping \citep{Bharadwaj:2000av, Bharadwaj:2003uh,bharadwaj04b}. This method allows the surveying of extensive cosmic volumes using current and upcoming radio telescopes \cite{Bharadwaj:2001vs, Bharadwaj:2003uh, Loeb:2008hg}.

In order to analyze the 21-cm intensity fluctuations, one of the choices is to compute the power spectrum of the intensity field. In the post-reionization era, the lack of complex reionization astrophysical processes in the intergalactic medium (IGM) makes the 21-cm power spectrum directly related to the underlying matter power spectrum \cite{Wyithe:2008mv} as well as underlying cosmology. Detecting Baryon Acoustic Oscillations (BAO) in the 21-cm power spectrum can provide precise constraints on the expansion rate of the universe, or on the equation of state for dark energy \cite{Chang:2007xk, Wyithe:2007rq, Seo:2009fq, Masui:2010mp}. Accurate measurement of the 21-cm power spectrum also holds promise for independently estimating various cosmological parameters \cite{Loeb:2008hg, Bharadwaj:2008yn, Obuljen:2017jiy}. Cross-correlating the 21-cm signal with other tracers of LSS, such as the Lyman~$\alpha$ forest \cite{Sarkar:2013yha, Carucci:2016yzq, Sarkar:2018vyh}, Lyman-break galaxies \cite{Villaescusa-Navarro:2014rra}, weak lensing \cite{Sarkar:2009zt}, and the integrated Sachs-Wolfe effect \cite{Sarkar:2008sz}, has been proposed as essential cosmological probes in the post-reionization era.

Several low-frequency instruments, including BINGO \cite{Abdalla:2021nyj, Costa:2021jsk} \footnote{https://bingotelescope.org/}, 
CHIME \citep{CHIME:2022dwe} \footnote{https://chime-experiment.ca/en}, 
Tianlai Project \citep{Chen:2012xu} \footnote{https://tianlai.bao.ac.cn/}, 
FAST \citep{FAST19} \footnote{https://fast.bao.ac.cn}, 
ASKAP \citep{ASKAP08} \footnote{https://www.atnf.csiro.au/projects/askap/index.html}, 
HIRAX \citep{HIRAX16}\footnote{https://hirax.ukzn.ac.za}, 
CHORD \citep{CHORD19} \footnote{https://www.chord-observatory.ca}, 
SKA \citep{SKA_SWG20} \footnote{https://www.skatelescope.org/}
are planned to measure the 21-cm signal at various redshifts of the post-reionization era. Recently, \cite{Paul:2023yrr} has reported the possible first detection of the 21-cm autocorrelation power spectrum at $z\sim 0.32$ and $0.44$ using the MeerKAT radio telescope. So it is expected that the signal will be detected at even higher redshifts in the post-reionization era.

The 21-cm signal is inherently weak, several orders of magnitude smaller than various astrophysical foregrounds \citep{ghosh11,ghosh12,switzer13}. The observed data from the present generation telescopes is mostly foreground and systematics dominated and we are yet to find an optimal method to isolate the signal from the contaminations. However, once this hurdle is over, we expect to detect the post-reionization 21-cm signal. We therefore need accurate models of the 21-cm signal to be able to estimate the amplitude and features of the power spectrum expected from observations. Not only that, we need accurate modelling to interpret the signal as well \citep{bull15,sarkar17}. 

Significant efforts have been devoted to modelling the post-reionization \HI\ distribution and the expected 21-cm signal. A number of analytic~\citep{Marin:2009aw,Padmanabhan:2016fgy,Penin:2017xyf}, semi-numerical~\citep{bagla10,khandai11,guha_sarkar12,sarkar16,Sarkar:2018gcb,Sarkar:2019nak,seehars16,Modi:2019ewx}, and fully numerical techniques~\citep{Dave:2013zja, Villaescusa-Navarro:2014cma, Villaescusa-Navarro:2018vsg, Rahmati:2012rg, Nelson:2018uso} have been implemented to simulate the \HI\ distribution and model the 21-cm signal statistics across the post-reionization redshifts. In this work, we rely on a semi-numerical simulation technique where the dark matter halos identified from N-body simulations are populated with \HI\ using an analytic prescription. For a comprehensive review of this technique, the reader is referred to the following references ~\citep{bagla10,khandai11,guha_sarkar12,sarkar16,seehars16}.

The standard cosmological model suggests that the initial density perturbations were largely Gaussian in nature \citep{Baumann:2009ds}. As the density fields evolve, non-Gaussianity is introduced due to the non-linearity in the fields induced by the flow of matter towards the high-density peaks at the biased locations in the Universe \citep{Fry:1983cj}. Also, a number of studies suggest that there may be a primordial non-Gaussianity component in the density fields \citep{Oppizzi:2017nfy}. Therefore, we need to consider primordial non-Gaussianity along with the induced non-Gaussianity in order to probe the complete information in the field. Whatever the scenario, ultimately, we have non-Gaussian matter distribution in the post-reionization era~\citep{gagrani17, Bharadwaj:2020wkc, Mazumdar:2020bkm, Mazumdar:2022ynd, joyce22, ivanov23, damico24, gill24}. The post-reionization 21-cm signal is expected to be non-Gaussian in nature as it directly follows the non-Gaussian matter distribution \citep{Sarkar:2019ojl}. Note that, unlike in the reionization 21-cm signal where the non-Gaussianity is majorly induced by the inhomogeneities in the radiation, heating, and ionization of the IGM gas, \citep{2018MNRAS.476.4007M, 2005MNRAS.358..968B} the non-Gaussianity in the post-reionization 21-cm signal is dominated by the density fields.

The two-point correlation or its Fourier conjugate, the power spectrum, is not an exhaustive statistic to capture the time-evolving non-Gaussianity in a field. For a Gaussian field, the power spectrum is sufficient to describe the field. However, for non-Gaussian fields, like the 21-cm signal, we need to think about ways to pick up information beyond the two-point correlation. One of the ways is to estimate higher-order correlations like the three- and four-point functions, or their Fourier conjugate, namely, bispectrum~\citep{Sarkar:2019ojl}, and trispectrum \cite{Shaw:2020xqj}. Calculating higher-order functions can be computationally expensive, although recent advancements show a significant speed-up in the computation~\citep{Shaw:2021pgy, Mazumdar:2022ynd, Nandi:2024cib}. Further, higher-order statistics can be difficult to interpret. Alternative ways include different summary statistics \citep{2010MNRAS.408.2373G, Yoshiura:2016nux}, strategies to segment the density field and calculate correlation, constructing marked fields, and studying their statistics that may contain the information of some higher-order moments~\citep{Aviles:2019fli,Karcher:2024twr,White:2016yhs,Satpathy:2019nvo}. For our current study, we mostly focus on the last alternative.

Several studies suggest and we shall also show in this work, the amplitude and shape of the \HI\ density power spectrum ($\Delta^2_{\rm HI}$) at small $k$ ($<1\;\mp$) do not evolve significantly across $1<z<5$ \citep{bagla10,khandai11,guha_sarkar12,sarkar16,Sarkar:2018gcb,Sarkar:2019nak,Modi:2019ewx,Villaescusa-Navarro:2018vsg}. On the other hand, the matter clustering grows significantly in this same $z$ range. Not only that, the \HI\ content of the dark matter halos, the \HI\ mass function, and the cosmological \HI\ (although, observationally, this has been found to change marginally over $z$, e.g. see \citep{Crighton:2015pza}), all evolve with $z$. The power spectrum mostly captures information from high-density regions. Therefore, it is expected to be sub-optimal for extracting information embedded in low-density regions and we shall not get any information on the \HI\ evolution at all densities by studying only $\Delta^2_{\rm HI}$.

In this work, we explore whether the marked power spectrum of the \HI\ density field can characterize clustering patterns in \HI\ distribution at different length scales. A mark of a field is a transformation of that field around each point. Mark usually depends on the properties of the considered field. Marked power spectrum is the Fourier conjugate of the two-point correlation function of the marked field. Since, in general, a mark is a non-linear function of the original field, the two-point marked correlation function not only contains the information of the original two-point correlation function but also at least some of the next higher-order correlation functions of the field. Here we study how the marks on the \HI\ density fields enable us to study the evolution and properties of the \HI\ distribution at various scales. The application of marked statistics to the $21$-cm signal here is the first of its kind for studying the post-reionization Universe. The marked power spectrum, however, has previously been used, for example, to detect the neutrino mass in the dark matter density distribution \cite{massara21} and information in the galaxy distribution \cite{massara23}. Estimating the marked power spectrum in this work is easy because all that it requires is a functional form of the local density as the mark. We just need to add this mark to the pipeline that computes the power spectrum while correlating the two marked points in order to estimate the marked power spectrum. Several previous studies have shown that certain non-linear transformations Gaussianize the density field up to an extent \cite{neyrinck09,neyrinck11b}. These transformations, therefore, turn the information inherent in the higher-order functions back to what we expect to obtain from the two-point correlations. This, thus, provides us with a way to improve the parameter constraints that are likely to be obtained from the two-point correlations.

The paper is organized as follows. In section \ref{sec:simulation} we present the \HI\ simulation. Section \ref{sec:formalism} describes the formalism of the marked correlation function and the marked power spectrum. In section \ref{sec:results} we discuss our analysis of the marked power spectra. Section \ref{sec:summary} presents a summary and discussion.

Throughout this paper, we have adopted the Planck+WP best-fit cosmological parameters $h=0.6704$, $\Omega_{\rm m}=0.3183$, $\Omega_{\Lambda}=0.6817$, $\Omega_{\rm b}h^2=0.022032$, $\sigma_8=0.8347$, $n_{\rm s}=0.9619$  from ref. \citep{planck14}.

%, which can be written as 
%\begin{equation}
    %P_{21}(k,z) = \bar{T}^2_{b}(z) P_{\rm HI}(k,z) = \bar{T}^2_{b}(z) b^2(k,z) P_{\rm m}(k,z)\,.
%\end{equation}
%Here $P_{\rm HI}$ is the \HI\ density power spectrum, $b(k,z)$ is the \HI\ bias, $P_{\rm m}$ is the matter power spectrum and $\bar{T}_{b}(z)$ is a factor to convert density into temperature \cite{Furlanetto:2006jb}. Note that, $\bar{T}_{b}(z)$ depends on a number of fundamental constants, Hubble expansion rate $H(z)$, and cosmological \HI\ density. 
% The formalization of correlations 
% in marked point processes was initially introduced by Ref.~[]. 
% These correlations have since found 
% applications in astrophysics, particularly in investigating the relationship between galaxy 
% clustering and various galaxy characteristics, including morphology, luminosity, color, and more. 
% A framework for describing marked correlation functions within the context of the halo model has 
% been established in Ref.~[]. An application of mark to study the modified gravity can be found in Ref.~[]. 
% Marks can also be utilised to study the signatures of neutrinos on large scale structures []. 

\section{Simulating the \HI\ distribution in the post-reionization era}
\label{sec:simulation}

We simulate the \HI\ distribution in the post-reionization era following three major steps. 
\subsection{\texorpdfstring{$\it N$}{nbody}-body simulation}
First, we simulate the dark matter distribution and velocity at redshifts in the range $z \in [1,5]$ in steps of $\Delta z = 1$, using a publicly available `dark-matter only' particle mesh (PM) $\it N$-body code\footnote{\url{https://github.com/rajeshmondal18/N- body}} \cite{mondal15}. The simulations start with $[1072]^3$ dark matter (DM) particles, each having a mass $1.09 \times 10^8\ M_{\odot}$, on $[2144]^3$ regular grids with a spatial resolution of $70$ kpc. This sets the comoving volume of our simulations to be $[150.08\ {\rm Mpc}]^3$. The initial conditions are set at $z = 125$ using the linear $\Lambda$CDM power spectrum \citep{eisenstein99} and the Zel’dovich approximation \citep{zeldovich70}. Starting from this, the $\it N$-body code then evolves the particle positions and velocities to desired redshifts in the range $z \in [1,5]$ and generates comoving dark matter snapshots.

\subsection{Finding FoF halos}
Next, we identify the collapsed dark-matter halos in the DM distribution by using a publicly available halo finder code\footnote{\url{https://github.com/rajeshmondal18/FoF- Halo- finder}} that employs the Friends-of-Friend (FoF) algorithm \cite{davis85}. Here, we set the linking length to $0.2$ in units of mean inter-particle spacing. We further impose a condition that a cluster of particles must at least have $10$ members to be identified as a halo, which sets the halo mass resolution to $ M_{\rm h, min} = 1.09 \times 10^9\ M_{\odot}$ in our simulations. The mass distribution of these FoF halos is also verified to be in good agreement with the theoretical halo mass function \cite{jenkins01, sheth02} in the range $10^9 \leq M_{\rm h} \leq 10^{13}\ M_{\odot}$.

\subsection{Simulating \HI\ using \texorpdfstring{$M_{\rm h}-M_{\rm HI}$}{mhmhi} relation}
Finally, we populate the identified FoF halos with \HI. We use an analytic prescription for populating the halos. Ref. \cite{bagla10} have proposed several schemes to connect \HI\ mass to the halo mass in the post-reionization era. In our work, we have considered the third scheme which connects \HI\ mass to halo mass as:
\begin{equation}
    M_{\rm HI}(M_{\rm h}) =
    \begin{cases}
      f_3 \frac{M_{\rm h}}{1+\frac{M_{\rm h}}{M_{\rm h,max}}}, & \text{if}\ M_{\rm h} \geq M_{\rm h, min} \\
      0, & \text{otherwise}\,.
    \end{cases}
    \label{eq:mh_mhi}
\end{equation}
$M_{\rm h, min}$ and $M_{\rm h, max}$ in this equation can be calculated using an approximate relation that connects the halo mass with the virial circular velocity ($v_{\rm circ}$) of the halo as:
\begin{equation}
    M_{\rm h} \simeq 10^{10} M_{\odot} \Bigg(\frac{v_{\rm circ}}{60\ {\rm km\ s^{-1}}}\Bigg)^3 \Bigg( \frac{1+z}{4} \Bigg)^{-\frac{3}{2}}. 
    \label{eq:vcirc}
\end{equation}

\begin{table}
\begin{center}    
\begin{tabular}{ |p{1cm}|p{1.5cm}|p{2cm}|p{2cm}| }
\hline
$z$ & $f_3$ & $M_{\rm h, min}$ ($10^{9}M_{\odot}$) & $M_{\rm h, max}$ ($10^{11}M_{\odot}$) \\
\hline
$1$ & $0.016$ & $3.53$ & $10.47$ \\
\hline
$2$ & $0.017$ & $1.92$ & $5.70$ \\
\hline
$3$ & $0.020$ & $1.25$ & $3.70$ \\
\hline
$4$ & $0.025$ & $0.89$ & $2.65$ \\
\hline
$5$ & $0.035$ & $0.68$ & $2.02$ \\
\hline
\end{tabular}
\end{center}
\caption{List of the values of the free parameter $f_3$ in $M_{\rm h}-M_{\rm HI}$ relation \ref{eq:mh_mhi}, $M_{\rm h,min}$ and $M_{\rm h,max}$ calculated using Equation \ref{eq:vcirc} at different redshifts.}
\label{tab:tab}
\end{table}

\begin{figure*}
    \centering
    \includegraphics[width=1\textwidth,angle=0]{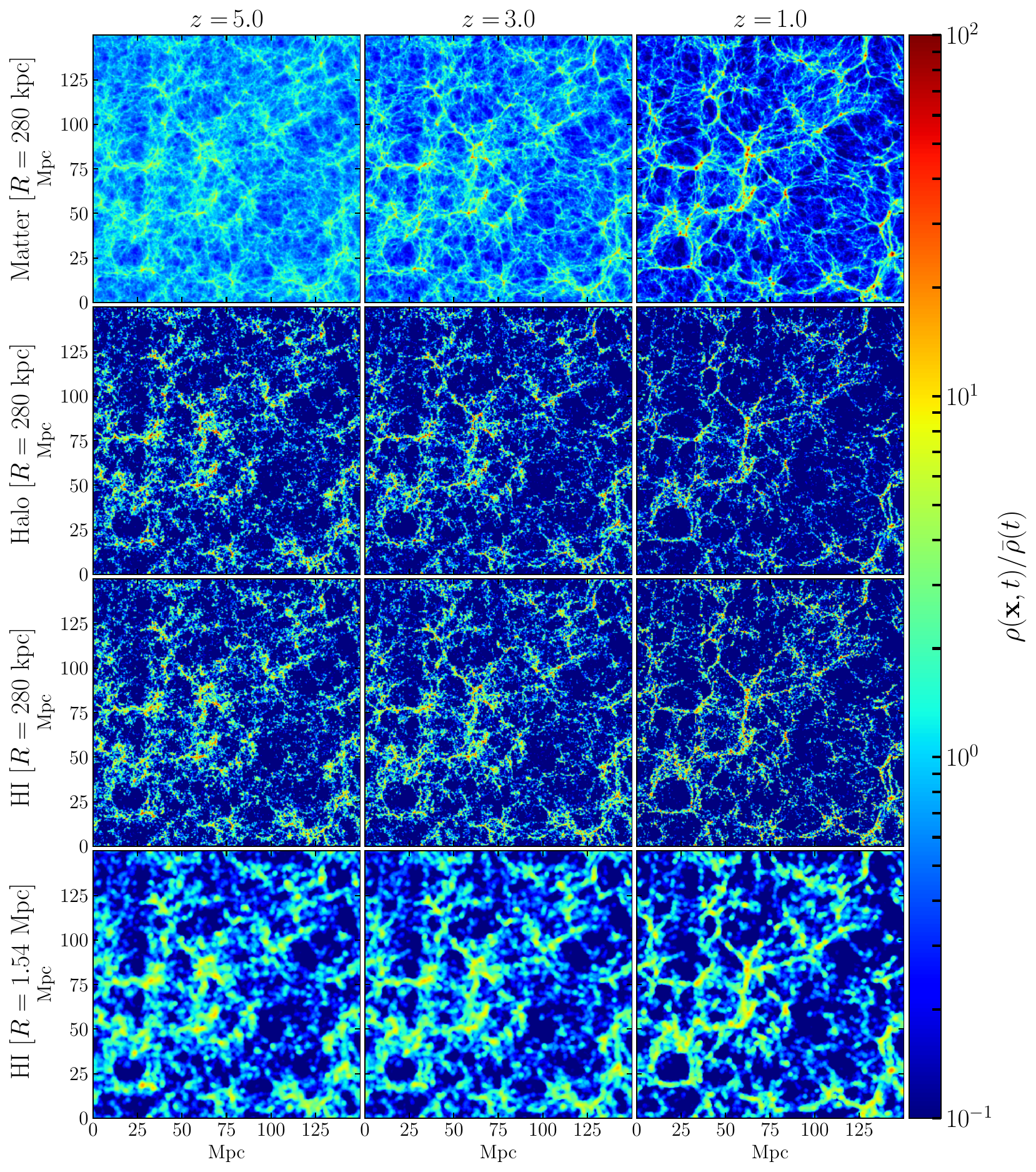}
    \caption{Visuals of matter (top first row), halo (second row) and \HI\ (third row) density distributions shown at three different redshifts $z=5$, $3$, and $1$, showing high-resolution distributions obtained after smoothing of this field using a spherical top hat filter for a fixed radius of $R = 280\, {\rm kpc}$ (grid cell size). Panels in the bottom row show the low-resolution \HI\ density distribution which is smoothed for $R = 1.54\, {\rm Mpc}$. For this, we first compute the densities on the grid points using the cloud-in-cell (CIC) interpolation scheme. We then plot the 2D slices showing the mean density on a plane after compressing a layer of thickness $5.6$ Mpc along a direction.}
    \label{fig:h1maps}
\end{figure*}

The $M_{\rm h}-M_{\rm HI}$ relation \ref{eq:mh_mhi} assumes that the \HI\ inside a halo survives only when the halo circular velocity exceeds a value of $\sim 30\ {\rm km\ s^{-1}}$, which determines the minimum threshold $M_{\rm h, min}$. The upper limit $M_{\rm h, max}$ is set by plugging $v_{\rm circ} \sim 200\ {\rm km\ s^{-1}}$, above which the \HI\ fraction in halos seems to decrease as suggested by simulations. The values of $M_{\rm h, min}$ and $M_{\rm h, max}$ at different $z$ are tabulated in Table \ref{tab:tab}. Note that, at $z=3.5$, $M_{\rm h, min} = 10^{9}\ M_{\odot}$ which is also the minimum mass of halos that we can resolve in our simulations. At $z>3.5$, \HI\ is allowed to reside inside halos with mass smaller than $M_{\rm h} < 10^{9}\ M_{\odot}$ which we cannot resolve given the mass resolution of our current setup. However, ref. \citep{sarkar16} discussed that ignoring the smaller halos has a very minor effect on the \HI\ power spectrum even at $z \sim 6$. 

This scheme is based on the following arguments. The majority of \HI\ in the post-reionization era is seen to be confined to dense clouds, known as Damped Lyman Alpha systems (DLAs), with column densities of $N_{\rm HI} \geq 2 \times 10^{20}\ {\rm cm}^{-2}$. These DLAs are observed in the absorption spectra of distant quasars (QSOs) \cite{wolfe05}, which also shows that very little \HI\ is left in the intergalactic medium at $z<5$. DLAs are hosted by galaxies that form inside the dark matter halos. Therefore, it is safe to assume that the total \HI\ in the post-reionization universe is locked inside the halos. We also assume that the \HI\ content of a halo solely depends on the halo mass $M_{\rm h}$, and within each halo \HI\ is distributed uniformly, rather than being contained within the substructures, like galaxies. A halo, however, needs to be above some critical threshold mass $M_{\rm h,min}$ to be able to self-shield the \HI\ content against the harsh ionizing UV background \citep{quinn96,dijkstra04,duffy12}. The \HI\ content of a halo above this threshold typically increases with the halo mass. Several low-redshift observations indicate that the \HI\ content drops for massive halos, typically above some threshold mass $M_{\rm h,max}$ \citep{pontzen08,dave13}. Therefore, it is expected that the halos having mass in the range $M_{\rm h,min} \le M_{\rm h} \le M_{\rm h,max}$ contain most of the \HI.

Scheme \ref{eq:mh_mhi} allows the halos with mass above $M_{\rm h,min}$ to host \HI. The \HI\ mass ($M_{\rm HI}$) inside a halo increases with the halo mass linearly until the halo mass attains a value $M_{\rm h, max}$ at which $M_{\rm HI}$ saturates to a value $M_{\rm HI} \approx (f_3) M_{\rm h, max}$. Therefore, for halos with $M_{\rm h} > M_{\rm h, max}$, the $M_{\rm HI}$ remains almost constant. The free parameter $f_3$ is adjusted by requiring that the cosmological \HI\ density parameter $\Omega_{\rm HI}(z)$ from simulations matches the observations \citep{rao06,noterdaeme12,crighton15}. The values of $f_3$ at different redshifts are listed in Table \ref{tab:tab}.

We have run our simulations for five statistically independent realizations. We use these realizations to estimate the mean and the standard deviation for all the results shown in this article. In Figure \ref{fig:h1maps}, we plot visualization of the density field $\rho({\bf{x}},z)/\bar{\rho}(z)$, the ratio of the density at ${\bf{x}}$ over the mean density, for matter (top first row), halo (second row), and the \HI\ (third and fourth row) for a single realization and at three different stages of cosmic evolution, $z=(5,3,1)$. The color bar in the log scale shows a clear evolution of the densities in the large range of their distributions. The matter and halo densities are smoothed using a spherical top hat filter of a fixed comoving radius of $R = 280\, {\rm kpc}$, while the \HI\ density is shown at two different resolutions obtained after smoothing of the distribution at comoving radii $R = 280\, {\rm kpc}$ and $1.54\, {\rm Mpc}$. Here, $R=280$ kpc is the, minimum possible scale (grid cell size) of smoothing in our simulation, whereas $R=1.54\, {\rm Mpc}$ represents the typical resolution of the intensity mapping experiments. 

A number of studies have shown that the scatter in the galaxy and halo luminosities impact the galaxy line intensity mapping signal and its statistics significantly over a wide $z$-range \citep{Li16,Yang22,Schaan21}. This scatter is also expected to impact the \HI\ distribution in the post-reionization era. We have not considered the effect of scatter in this paper.

\section{Formalism of marked statistics}
\label{sec:formalism}

\subsection{The marked correlation function}
Two point marked correlation function (MCF) in configuration space is defined as \cite{sheth05a,white16}
\begin{equation}
    \mathcal{M}(x) = \frac{1}{n(x)\bar{m}^2}\sum_{ij}m_i m_j = \frac{1+W}{1+\xi}\,,
    \label{eq:mcf}
\end{equation}
where $n(x)$ is the number of pairs at separation $x$ in a field, the summation over which is needed to get the value of $\mathcal{M}$. $m_i$ is the mark at $i_{\rm th}$ point with a mean $\bar{m}$ over the entire sample. The second equality is set to define $W$ which is normalized by the two-point correlation function ($\xi$) to emphasize the $\mathcal{M}$ to be $1$ at large scales. $\mathcal{M}$ is easy to compute if one already has a pipeline to compute the standard two-point correlation function. It only requires a simple modification to the correlation function calculations.

The correlations of the marked point processes for the first time were introduced by \cite{stoyan83}. A wide range of astrophysical and cosmological applications of the MCF have subsequently been made. This, for example, has been used to study how the bars and bulges in disc galaxies depend on the environment \citep{skibba12}. This also offers a way of testing the connection between galaxy clustering and galaxy properties such as luminosity, morphology, color, star formation rate, stellar mass, etc. \citep{beisbart2000,beisbart02,sheth05a,skibba06}. Furthermore, the marked statistics are able to trace the alignment of dark matter halos in the distribution of the dark-matter particles \citep{beisbart02}, and to probe the dependence of clustering of these halos on merger history \citep{gottlober02}. A halo model description of the MCF was developed in \citep{sheth05b}. The MCF has also been used to break degeneracies in different halo occupation distribution (HOD) models \citep{white09}. In the past decade, marked statistics have extensively been used to demonstrate how different modified gravity (MG) models are distinguished from general relativity (GR), due to the strong dependence of the tracer on the environment in MG models \citep{zhao11,winther12,lombriser15, white16,shi17,armijo18,valogiannis18,satapathy19,hernandez19,armijo23}.

\subsection{The choices of the mark}
In this work, we use the mark as a function of local density ($\rho_{R}$) formulated by \cite{white16}, and given as
\begin{equation}
    m(\textbf{x},t; R,\rho_{\ast}, p) = \Bigg[\frac{\rho_{\ast}+1}{\rho_{\ast}+\rho_R(\textbf{x},t)}\Bigg]^p \, ,
    \label{eq:m_mw16}
\end{equation}
where, the free parameters $R,\ \rho_{\ast},\ {\rm and}\ p$ are known as the mark parameters. Here, both $\rho_{R}$ and $\rho_{\ast}$ are in unit of mean density $\bar{\rho}$. The parameter $\rho_{\ast}$ reveals how strongly the mark is sensitive to the local density around each point. The exponent $p$ introduces an additional contrast between two points in different environments by weighting these points differently. Note that the choice of the functional form of the mark, as well as the values of the free parameters, is arbitrary, but needs to be chosen according to the properties of the given field and the information that we are interested in extracting from that field.

The idea of using the mark as a function of local density was first proposed by \cite{white09} to constrain HOD models in order to study which galaxies reside inside which halos. The mark $m$ was introduced to deal with cases where we have a smoothed field of the observable. $m$ was designed as a function of $\rho_{R}$, smoothed over some radius $R$ using filters like spherical top-hat, for extracting the information around the objects without knowing their own properties. The observations of the redshifted 21-cm signal using the intensity mapping techniques are expected to produce low-resolution maps of the 21-cm intensity fluctuations. Since the \HI\ density field is proportional to the 21-cm intensities, we therefore can use the \HI\ density to define the mark as given by equation~\ref{eq:m_mw16}. We apply this mark at all the \HI\ redshift snapshots and create marked \HI\ fields for studying the \HI\ clustering and evolution with time in the post-reionization era. We have also considered a different functional form for the mark proposed in \cite{white09} and defined in equation~\ref{eq:m_mw09}. We found that this mark is not very useful for our purpose and we briefly discuss the associated results in Section~\ref{app:mark_W09}. From here onwards, we shall focus mainly on the mark defined in equation~\ref{eq:m_mw16}.

\begin{figure*}
    \centering
    \includegraphics[width=1\textwidth,angle=0]{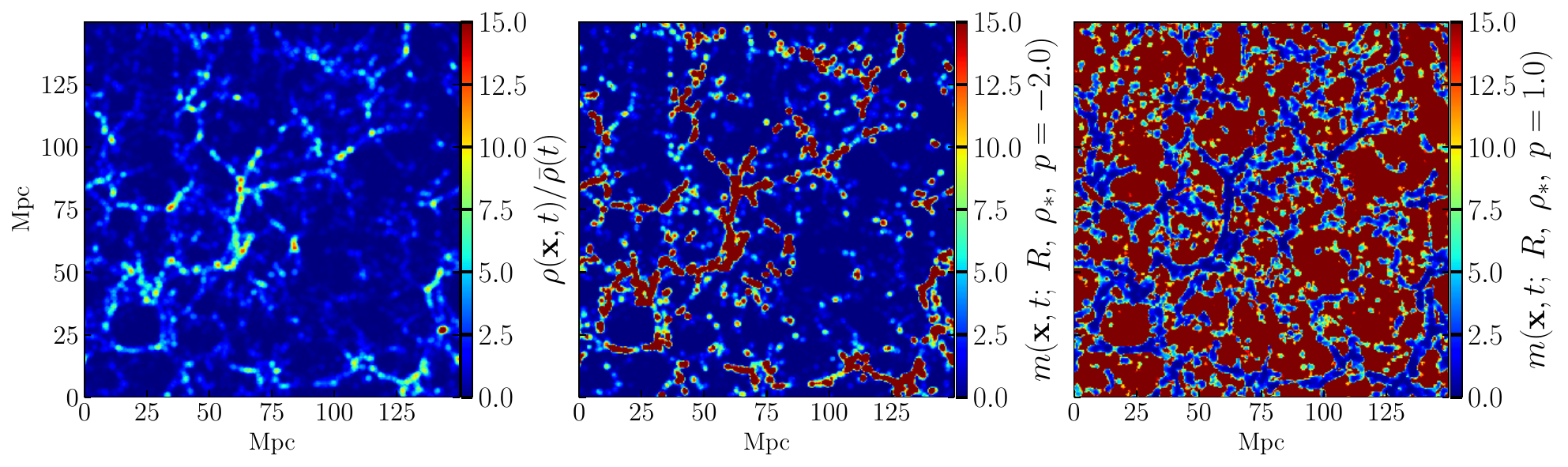}
    \caption{\textbf{Left panel:} \HI\ density field at redshift $z =1$ which is smoothed for a radius $R = 1.54$ Mpc. \textbf{Middle and Right panels:} Marked \HI\ density fields at $z =1$, with parameters $R = 1.54$ Mpc, $\rho_{\ast} =0.01$ and for $p_{\rm hd}$ (middle panel) and $p_{\rm ld1}$ (right panel).}
    \label{fig:h1_mark}
\end{figure*}

\begin{table}
\begin{center}    
\begin{tabular}{ |p{1.5cm}|p{2.5cm}|p{2cm}|  }
\hline
$\rho_{\ast}/\bar{\rho}$& $p$ &$R$ (Mpc) \\
\hline
\multirow{2}{4em}{$0.01$} & \multirow{2}{4em}{$-2.0$\,($p_{\rm hd}$)} & $0.28$ \\ 
& & $1.54$ \\
\hline
\multirow{2}{4em}{$0.01$} & \multirow{2}{4em}{$1.0$\,($p_{\rm ld1}$)} & $0.28$ \\ 
& & $1.54$ \\
\hline
\multirow{2}{4em}{$0.01$} & \multirow{2}{4em}{$2.0$\,($p_{\rm ld2}$)} & $0.28$ \\ 
& & $1.54$ \\
\hline
\end{tabular}
\end{center}
\caption{List of the values of the free parameters of the mark $m$ (eq. \ref{eq:m_mw16}) considered in this paper. We designate positive $p$ as $p_{\rm ld1}$, $p_{\rm ld2}$ for $p=1,2$. On the other hand, we designate negative $p$ as $p_{\rm hd}$ for $p=-2.0$, as we have considered only one negative value.}
\label{tab:tab1}
\end{table}

For $\rho_{\ast} \rightarrow 0$, the mark $m$ is informative as it scales as  $m \approx [1/\rho_R]^p$. This suggests that, for positive $p$ (which we designate as $p_{\rm ld1}$, $p_{\rm ld2},\dots$ for $p=1,2,\dots$), $m$ gives more weight to underdensities and therefore is sensitive to the information in the underdense regions. On the other hand, for negative $p$ (which we refer to as $p_{\rm hd}$ for $p=-2.0$, as we have considered only one negative value), $m$ applies more weight to overdensities and probes features from the overdense regions. Table \ref{tab:tab1} lists the values of free parameters used in this paper.

Figure~\ref{fig:h1_mark} shows a visual comparison between the standard \HI\ density field (left panel) together with the two marked density fields (middle panel for $p_{\rm hd}$ and right panel for $p_{\rm ld1}$). Comparing the left and middle panels, we see that the high-density regions in the middle panel are very prominent as compared to the left panel which is exactly what we expect for the mark with $p_{\rm hd}$. On the other hand, doing the same comparison between the left and right panels, we find that the underdense regions are enhanced in the right panel for $p_{\rm ld1}$. Since for the standard density field, the power spectrum mostly picks contributions from the high-density regions, we expect the marked power spectrum for $p_{\rm hd}$ to exhibit similar features as in the power spectrum for the standard density field. On the other hand, the marked power spectrum with $p_{\rm ld1}$ is expected to show different behaviour and contain additional information as compared to the standard density power spectrum.

\begin{figure*}
    \centering
    \includegraphics[width=1\textwidth,angle=0]{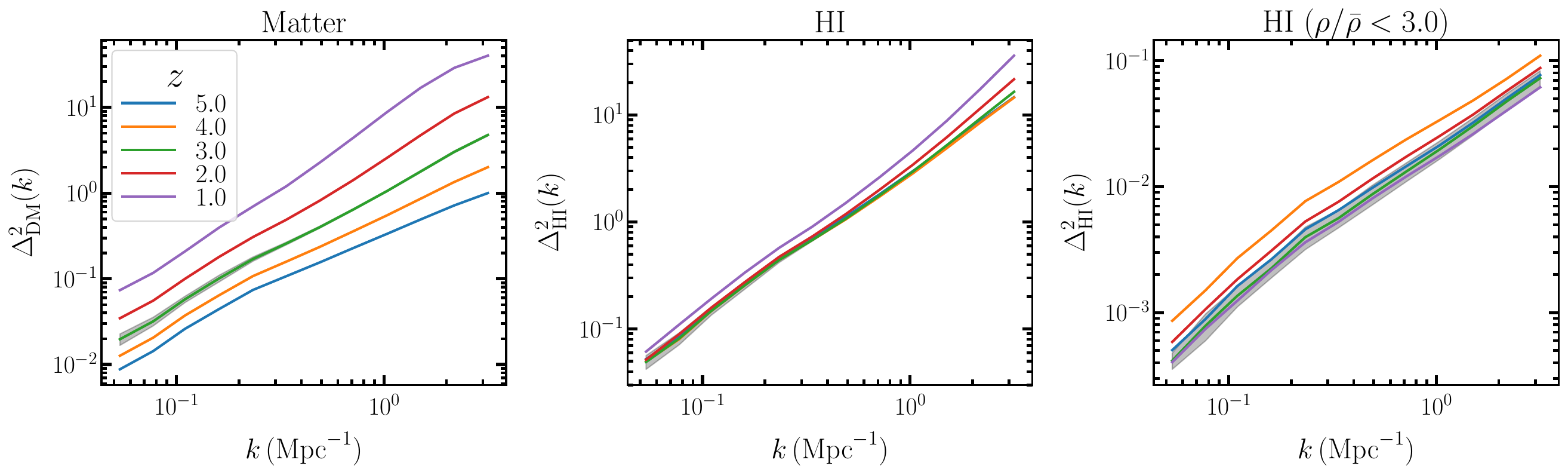}
    \caption{The dimensionless power spectrum of the matter (left panel), the \HI\ (middle panel), and the \HI\ with density values $[\rho/\bar{\rho}] < 3.0$ (right panel). The distribution is smoothed for a fixed radius of $R = 280\, {\rm kpc}$. The power spectra are demonstrated as a function of $k$-mode at five different redshifts in the range $z \sim 1$--$5$. The shaded regions in each panel represent $\pm1\sigma$ spread around the mean value at $z = 3$.}
    \label{fig:matter_h1_pk}
\end{figure*}

\section{Results} 
\label{sec:results}
Figure~\ref{fig:h1maps} shows the features that we can visually infer by looking at the dark matter, halo, and \HI\ density snapshots. The halo field closely follows the dark matter density peaks. Also, the \HI\ distribution traces the halo distribution at all $z$, which is expected as we have populated the halos with \HI. The dark matter clustering grows with time, as we go to lower $z$. As a result, the halos grow in mass and number with decreasing $z$, which results in the growth of the density contrast. According to the \HI\ population model in Equation~\ref{eq:mh_mhi}, the \HI\ mass inside the halos exceeding a certain cut-off mass $M_{\rm h,max}$ remains almost fixed. As a result, although the halos grow in mass with time and the halo density contrast increases, halos that surpass the upper threshold do not show any enhancement in density contrast for \HI. On the other hand, some halos that were devoid of any \HI\ at higher redshifts and cross the minimum threshold $M_{\rm h,min}$ at any lower $z$, appear as new overdense regions in \HI\ distribution and contribute to the evolution of the \HI\ field. However, it is difficult to locate these changes visually from Figure~\ref{fig:h1maps}. Thus, the \HI\ distribution is biased against the dark matter distribution. Visually, the \HI\ distribution, on the contrary of matter and halo distribution, evolves only weakly with $z$ due to this \HI\ density bias.

In Figure~\ref{fig:matter_h1_pk}, we show the evolution of the spherically averaged dark-matter power spectrum ($\Delta^2_{\rm DM}(k)$ in the left panel), along with the \HI\ density power spectrum ($\Delta^2_{\rm HI}(k)$ in the middle panel) in the redshift range $z\in[{1,5}]$. In order to understand the weak evolution of $\Delta^2_{\rm HI}(k)$ with $z$, in the right panel, we have plotted the \HI\ power spectrum for low- and intermediate-density regions where $[\rho/\bar{\rho}]<3$ (which we refer to as $\Delta^2_{\rm HI(LID)}(k)$). Here, we have normalized the power spectrum to the dimensionless quantity $\Delta^2(k) = k^3P(k)/2\pi^2$. We use five realizations of the simulations, each providing a statistically independent estimate of the power spectrum, to compute the mean and standard deviations. The lines in the figures show the mean at different redshifts and a shaded region around the mean shows $\pm1\sigma$ standard deviation. Note that the power spectrum and marked power spectrum have been estimated using a grid $4$ times courser than the one we used for the {$\it N$}-body simulations. In the following sections, we also restrict our analysis to $k$ modes in the range $0.05 \lesssim k \lesssim 1.0$ Mpc$^{-1}$ and smoothing radii in the range $0.28 \lesssim R \lesssim 5.0$ Mpc.

Considering the left panel of Figure~\ref{fig:matter_h1_pk}, we find that $\Delta^2_{\rm DM}(k)$, starting from $z=5$, grows considerably with decreasing $z$. This is due to the growth in dark matter clustering with time owing to gravitational instability. At $k<1\mpci$, $\Delta^2_{\rm DM}(k)$ shows nearly a power law behaviour with $k$. Considering the middle panel, we find that $\Delta^2_{\rm HI}(k)$ also shows a similar power law behaviour with $k$. However, contrary to $\Delta^2_{\rm DM}(k)$, $\Delta^2_{\rm HI}(k)$ shows very little evolution over $z$. In the right panel, we see that $\Delta^2_{\rm HI(LID)}(k)$ exhibits considerable variations with $z$ indicating that the clustering at low and intermediate densities evolves with $z$. However, comparing the right and the middle panels, we see that the amplitude of $\Delta^2_{\rm HI(LID)}(k)$ is nearly two orders of magnitude smaller than that of the total \HI\ power spectrum. Therefore, even though these regions evolve, their overall contribution is negligible in comparison to the high-density regions. We have already discussed in the context of Figure~\ref{fig:h1maps} that the high-density \HI\ regions do not show considerable enhancement in clustering with $z$ due to the \HI\ density bias. This is the main reason for the weak redshift evolution of $\Delta^2_{\rm HI}(k)$ over time in the middle panel. Even though there is a significant evolution in the low-density \HI\ regions, the same is not captured well by the power spectrum.

\begin{figure*}
    \centering
    \includegraphics[width=1\textwidth,angle=0]{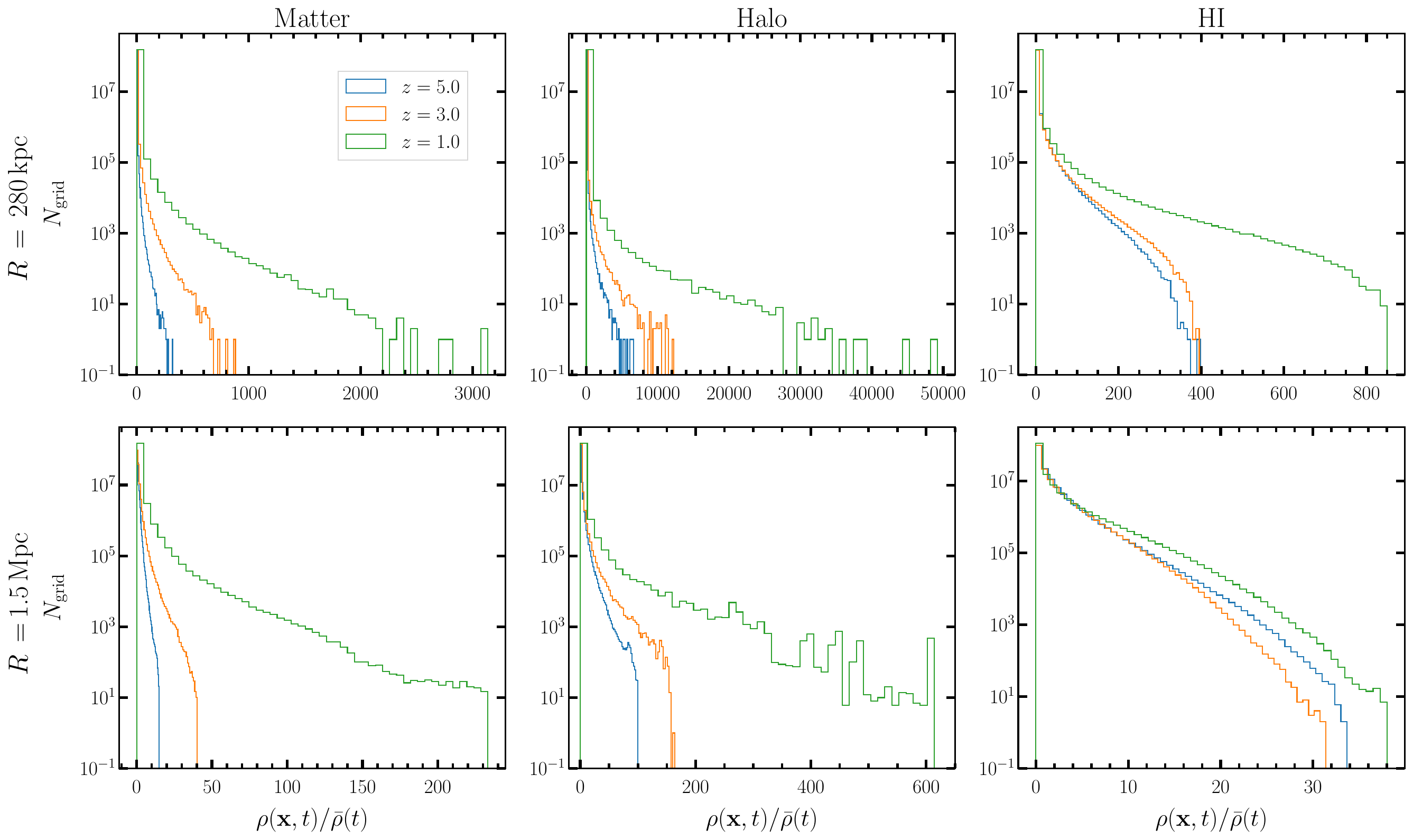}
    \caption{Histogram of the density distributions of the matter (left column), halo (middle column), and \HI\ (right column) on the grids in the simulation volume at three different redshifts, $z=(1$, $3$, $5$). The upper and lower panels are shown respectively for two different $R$ values; $R = 280\, {\rm kpc}$ and $1.54\, {\rm Mpc}$. The different histograms in each panel show how the density is being evolved over the cosmic time in the post-reionization era.}
    \label{fig:histogram}
\end{figure*}

In order to understand the evolution of \HI\ densities relative to the densities of matter and halos, we show the histogram of the quantity $\rho({\bf x})/\bar{\rho}$ for the three fields at three redshifts in Figure~\ref{fig:histogram}. Note again that these densities are computed on the grids. We have used two different smoothing radii $R = 280\, {\rm kpc}$ and $R = 1.54$ Mpc for computing the densities. Considering all the panels, we see that the evolution of the histogram with $z$ is weaker for \HI\ compared to both matter and halos. This weaker evolution in \HI\ is substantial to consider for studying the \HI\ clustering using the intensity mapping experiment. This conclusion remains the same for the two different aforementioned smoothing scales.

\begin{figure*}
    \centering
    \includegraphics[width=1\textwidth,angle=0]{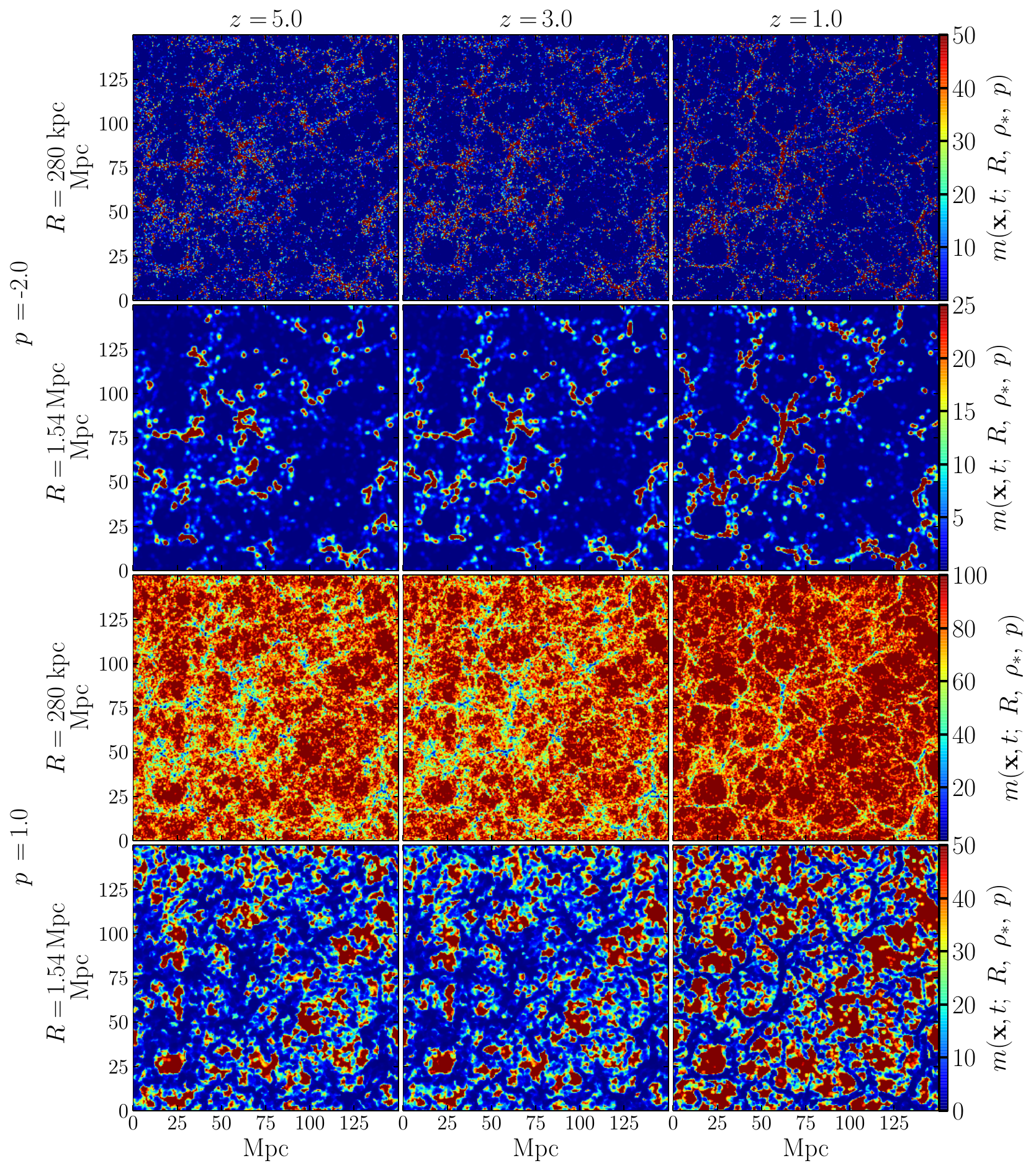}
    \caption{Marked \HI\ field ($m$) at different redshifts (decreasing from left to right) and for different mark parameters (varying in the vertical direction). Here, the value of the parameter $\rho_{\ast}$ is $0.01$. The top two rows show the time evolution in $m$ for two different radii $R = 280$ kpc and $R = 1.54$ Mpc and for a fixed parameter $p_{\rm hd}$. The bottom two rows show the same but now for $p_{\rm ld1}$.}
    \label{fig:m1_MW16}
\end{figure*}

\begin{figure*}
    \centering
    \includegraphics[width=0.76\textwidth,angle=0]{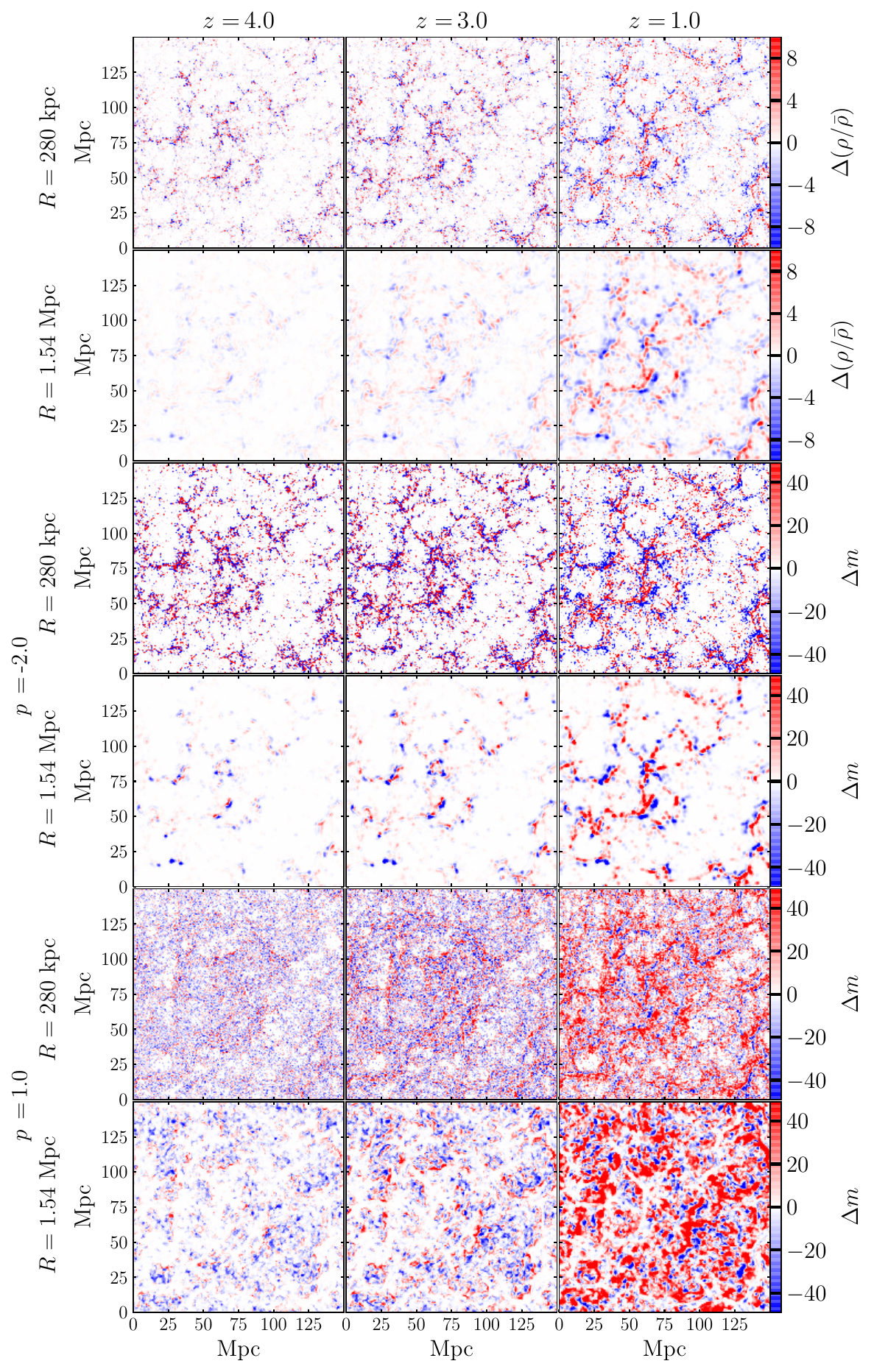}
    \caption{\textbf{Top two rows:} The quantity $\Delta (\rho/\bar{\rho})({\bf x},z) = (\rho/\bar{\rho})({\bf x},z)-(\rho/\bar{\rho})({\bf x},z=5)$ and \textbf{bottom four rows:} $\Delta m({\bf x},z) = m({\bf x},z)-m({\bf x},z=5)$ that alternatively visualizes the evolution found in Figure \ref{fig:m1_MW16}. For a particular field, this represents the difference in the field at any $z$ and $z=5$ for two $R$ values i.e., $R=280$ kpc and $1.54$ Mpc. The two middle and two bottom rows respectively show the difference for the marked field $m$ with $p_{\rm hd}$ and $p_{\rm ld1}$.}
    \label{fig:map_subtract}
\end{figure*}

The evolution of the \HI\ field can be visualized alternatively by plotting the quantity $\Delta (\rho/\bar{\rho})({\bf x},z) = (\rho/\bar{\rho})({\bf x},z)-(\rho/\bar{\rho})({\bf x},z=5)$, where $(\rho/\bar{\rho})({\bf x},z)$ is defined at $({\bf x},z)$ and $(\rho/\bar{\rho})({\bf x},z=5)$ is the same field but defined at $z=5$. The top two rows of Figure~\ref{fig:map_subtract} show this quantity at three redshifts $z=4, 3$ and $1$ and for two different $R$. $\Delta (\rho/\bar{\rho})$ here captures the change in the \HI\ density at a point due to gravitational clustering relative to $z=5$. In other words, $\Delta (\rho/\bar{\rho})$ captures the transfer of \HI\ from low-density regions to high-density regions due to the growth of structures. We see that for each $R$, $\Delta (\rho/\bar{\rho})$ changes considerably with $z$. $\Delta (\rho/\bar{\rho})$ is low at higher $z$ and high at lower $z$ due to the evolution of \HI\ with time. Also, this change depends on $R$, which is larger for $R=280$ kpc compared to that for $R=1.54$ Mpc. Thus, the \HI\ density shows evolution, and this evolution is not captured well enough by the normal power spectrum.

Before applying the mark, it is essential to decide on the values of the mark parameters, $R,\ \rho_{\ast}, {\rm and}\, p$, as the various combinations of these determine the different features that can be extracted from the density fields. We consider a total of six combinations of the mark parameters listed in Table~\ref{tab:tab1}. In general, the parameter $R$ can take any arbitrary value, but in practice, it should be chosen to match the resolution of a particular experiment. The choice of parameter $\rho_{\ast}$ is also very crucial as this is added to the density at each grid, shifting the mean value of the field. Further, in practice, there may be inherent noise in the density field estimation and, therefore, we choose smaller $p$ values, $|p| \leq 2.0$, to prevent the noise from getting amplified. We recapitulate once more that with $p_{\rm ld1}$ or $p_{\rm ld2}$, the mark $m$ gives more weight to the underdensities. Whereas, $p_{\rm hd}$ applies more weight to the overdensities. 

%\subsection{Marked \HI\ power spectra (MPS) for a fixed \texorpdfstring{$\rho_{\ast}$}{rho} and different \texorpdfstring{$p$}{p} and \texorpdfstring{$R$}{R}}

\subsection{The Fourier mode (\texorpdfstring{$k$}{k}) dependence of the marked \HI\ power spectra (MPS)}
\label{result:ps_m1}

In Figure~\ref{fig:m1_MW16}, we show the marked fields for $p_{\rm hd}$ and $p_{\rm ld1}$, at three redshifts and for two $R$ values, $280$ kpc and $1.54$ Mpc. Note that unless otherwise stated, $\rho_{\ast}$ is kept fixed at $0.01$ for all the figures. The impact of considering different values of $\rho_{\ast}$ on the results has also been discussed in the section \ref{app:ps_m1_diif_rho}. Keeping Figures~\ref{fig:h1maps} and \ref{fig:m1_MW16} side by side, we can broadly say that the over-dense regions in Figure~\ref{fig:h1maps} (last two rows) are enhanced for $p_{\rm hd}$ and are suppressed for $p_{\rm ld1}$. For example, the prominent overdensity at box location $\sim (75, 75)$ Mpc is enhanced for $p_{\rm hd}$ and is suppressed for $p_{\rm ld1}$. Further, the mark with $p_{\rm ld2}$ only enhances the contrast between high and low-density regions, giving more weight to the overdensities.

\subsubsection{MPS for \texorpdfstring{$p_{\rm hd}$}{p-2}}

In order to visualize the evolution found in Figure~\ref{fig:m1_MW16} in an alternative way, we have plotted the quantity $\Delta m({\bf x},z) = m({\bf x},z)-m({\bf x},z=5)$ in  Figure~\ref{fig:map_subtract} at three redshifts. The middle and the last two rows in this figure respectively show $\Delta m$ with $p_{\rm hd}$ and $p_{\rm ld1}$, calculated using two $R$ values. Considering the two middle rows, we find that the basic web structure is similar to that of the top two rows which represent $\rho/\bar{\rho}$, only the amplitude of the overdensities increased. Also, the redshift evolution of the high-density peaks looks more prominent than in $\rho/\bar{\rho}$.

\begin{figure*}
    \centering
    \includegraphics[width=1\textwidth,angle=0]{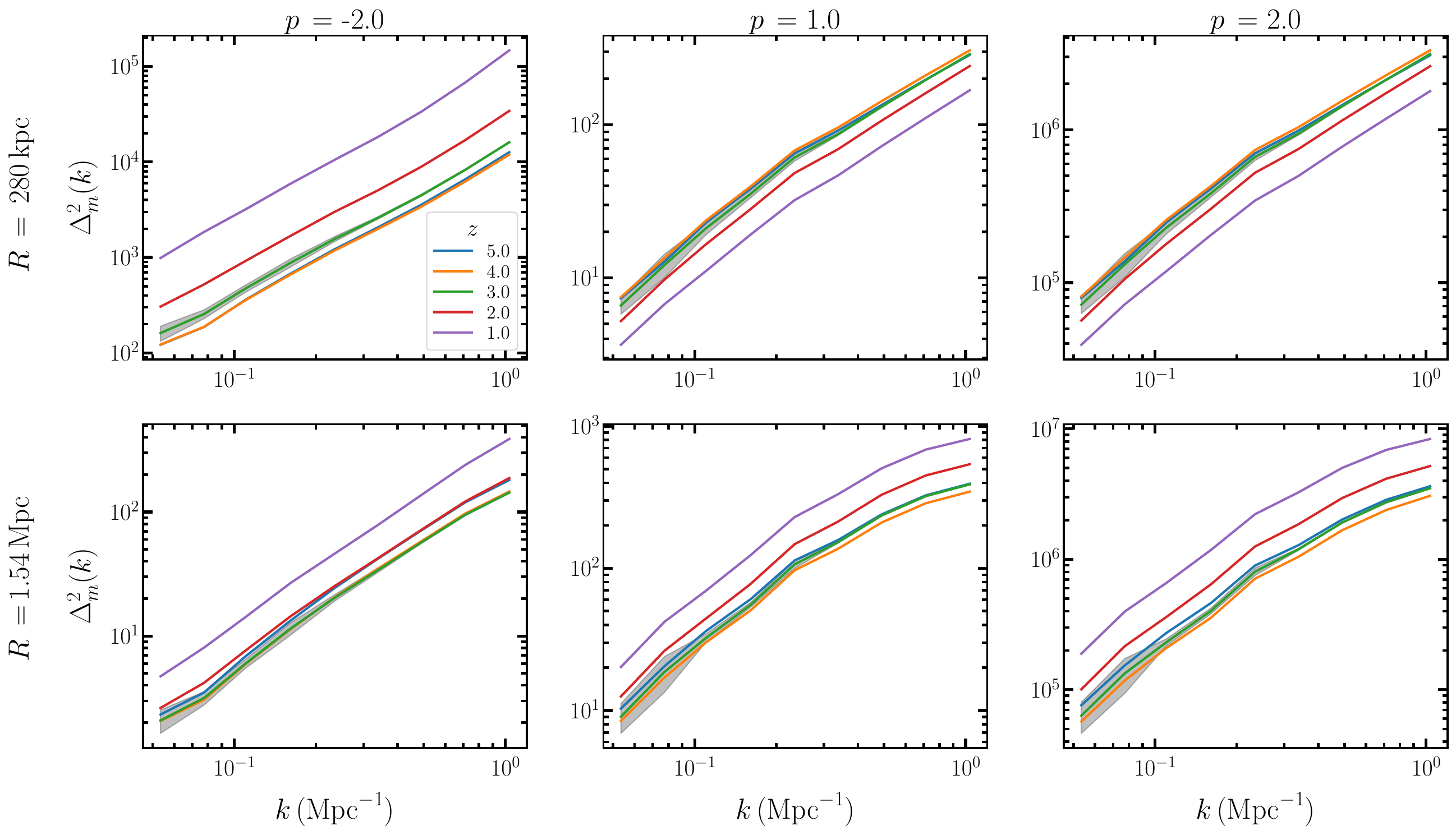}
    \caption{Marked \HI\ power spectra ($\Delta^2_{m}$) as a function of $k$ mode at five different redshifts. Here, we consider three different $p$ values and two different $R$ values. The upper panels are at $R=280$ kpc and the lower panels are at $R=1.54$ Mpc. The three columns are at three different values of the parameter $p$ which are $p_{\rm hd}$ (left-column), $p_{\rm ld1}$ (middle-column), and $p_{\rm ld2}$ (right-column). The shaded regions in each panel represent $\pm1\sigma$ spread around the mean at $z = 3$.}
    \label{fig:pk_MW16}
\end{figure*}

We shall now use the MPS to statistically verify whether this intuition is correct. In Figure~\ref{fig:pk_MW16}, we show the MPS ($\Delta^2_{m}$) at various redshifts for different values of the mark parameters $p$ and $R$. Considering the left column, which shows $\Delta^2_{m}$ for $p_{\rm hd}$ with two $R$ values $280\;{\rm kpc}$ and $1.54\;{\rm Mpc}$ respectively, we find that the redshift evolution of $\Delta^2_{m}$ resembles that of $\Delta^2_{\rm HI}$, as expected. $\Delta^2_{m}$ increases with decreasing $z$, shows a power-law behaviour with $k$ and the separation between $\Delta^2_{m}$ vs $k$ lines at two successive $z$ values (here $\Delta z=1$) increases with decreasing $z$. This indicates that the amplitude of \HI\ overdensities grows with time, and the cosmic web structures along the overdense regions become sharper. The amplitude of $\Delta^2_{m}$, however, is dependent on $R$ and, at a fixed $z$, we find a smaller amplitude for $R=1.54\;{\rm Mpc}$. Also, for a fixed $\Delta z$, the separation between the lines decreases as compared to that for $R=280$ kpc. All this is expected as with increasing $R$, the peaks in $m$ get smoothed out, which results in a reduced amplitude in the MPS. We shall investigate the MPS for a wide range of $R$ in section \ref{result:ps_m_smr} that can be considered to cover a wide range of $21$-cm observations of the post-reionization era with different telescopes. 

%Overall, Figure~\ref{fig:map_subtract} again suggests that the mark with $p_{\rm ld1}$ can track the evolution of the underdense regions, whereas, the mark with $p_{\rm hd}$ tracks the evolution of the overdense regions.

\subsubsection{MPS for \texorpdfstring{$p_{\rm ld1}$}{p1}}

Considering the two bottom rows in Figure \ref{fig:map_subtract}, which show the marked field with $p_{\rm ld1}$, we find that the high-density peaks as in the top four rows are missing here. This is more clearly visible for $R=1.54$ Mpc panels. The redshift evolution in the bottom two panels is mostly due to the low-density regions, exhibiting substantial evolution with redshift. Even though for different $R$, the marked fields visually look different, qualitatively the evolution looks similar. 

In order to connect this evolution with the MPS, we now consider the middle and the right columns in Figure \ref{fig:pk_MW16}. We find that for a fixed $R$, the $z$ evolution and $k$ dependence of $\Delta^2_{m}$ is similar for the two $p$ values, only the amplitude is higher for the higher $p$. Here also, the separation between $\Delta^2_{m}$ vs $R$ lines at $\Delta z=1$ increases with decreasing $z$. However, the trend in the $z$ evolution changes, as we vary $R$, keeping the $p$ value fixed. For $R=280\;{\rm kpc}$, the amplitude of $\Delta^2_{m}$ decreases with decreasing $z$, whereas the opposite happens for $R=1.54\;{\rm Mpc}$. The shape of $\Delta^2_{m}$, (the $k$ dependence), also changes with $R$. This behaviour can be understood as follows. Considering the third row of Figure~\ref{fig:m1_MW16}, which shows the marked field with $p_{\rm ld1}$ and $R=280\;{\rm kpc}$, we find that the cosmic web structures along which we see lower values of $m$ (which correspond to the overdensities in the \HI\, distribution) are sharp and prominent. The high $m$ (or underdensity) regions are bounded by these webs. $m$ attaining the peak values in these regions occupy a larger fraction of the simulation volume than the other $m$ regions and thus they act as a nearly constant background. At the highest redshift $z=5$, the size of these regions was smaller. However, as the redshift decreases, matter flows from low- to high-density regions, making the cosmic web structures sharper and the $m$ regions with peak values wider. Since the regions with peaks of $m$ act as background here, the power spectrum is dominated by the low $m$ regions where the amplitude decreases as matter flows in with decreasing $z$. This can also be seen in the fifth row of Figure~\ref{fig:map_subtract} where $\Delta m$ is most prominent in the low $m$ regions. Due to this, $\Delta^2_{m}$ decreases with decreasing $z$ for $R=280\;{\rm kpc}$.

Something remarkable happens in the marked field when we use $R=1.54\;{\rm Mpc}$, as shown in the last row of Figure~\ref{fig:m1_MW16}. Due to the smoothing over a large scale, the low $m$ regions spread out taking away portions from the high $m$ regions. These low $m$ regions are no longer sharp, rather they occupy most of the volume, squeezing the boundaries of the high $m$ regions, and making them appear like small isolated islands. These high $m$ islands are smaller and fewer in number at the highest redshift $z=5$. On the other hand, the low $m$ regions occupying a larger fraction of the simulation volume than the high $m$ regions, appear as a background on top of which the islands are painted. The last row of Figure~\ref{fig:map_subtract} shows that $\Delta m$ is nearly zero in the low $m$ regions, suggesting that the flow of matter from low to high density does not get captured in terms of no change in the density along the cosmic web structures is seen when the field is smoothed over a sufficiently large scale. On the other hand, $\Delta m$ is in non-zero at the high $m$ regions, indicating that the power spectrum is dominated by these regions. As the redshift decreases, the size and number of the high $m$ regions increase, which increases $\Delta m$ with decreasing $z$. Due to this, $\Delta^2_{m}$ increases with decreasing $z$ for $R=1.54\;{\rm kpc}$.

\begin{figure*}
    \centering
    \includegraphics[width=1\textwidth,angle=0]{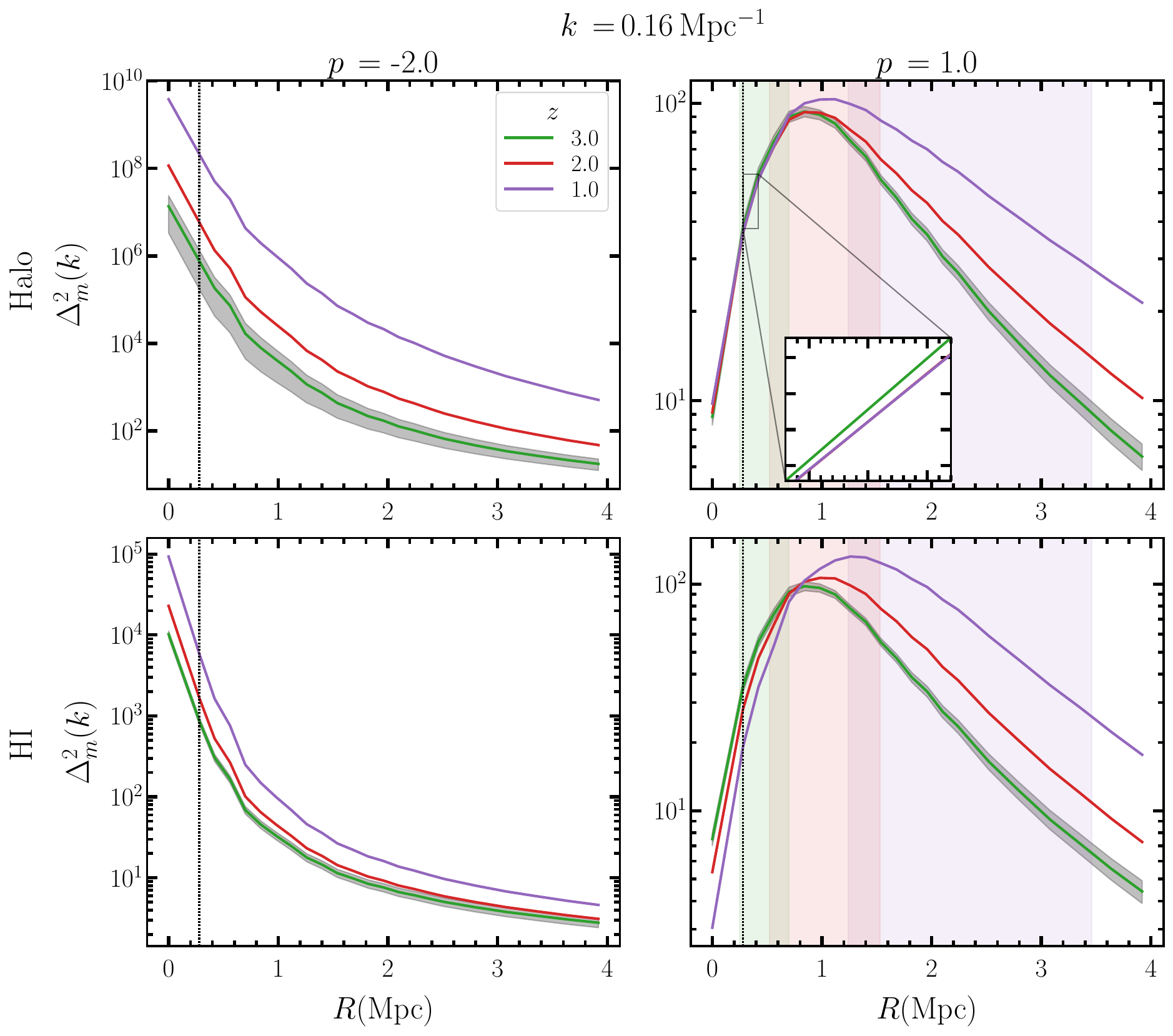}
    \caption{$\Delta^2_{m}$ for halo (upper panels) and \HI\ (lower panels) fields as a function of $R$ for $k = 0.16 \mp$ and at three different redshifts. We consider the mark parameters $\rho_{\ast}=0.01$ and two different $p$ values, $p_{\rm hd}$ (left panels) and $p_{\rm ld1}$ (right panels). The shaded regions around the line at $z=3$, in each panel of the right column, represent $\pm1\sigma$ confidence interval around the mean value. Each vertical shaded region is the range between two different $R$ values i.e. $R_{\rm ta}$ (right end) and $R_{\rm c}$ (left end) that respectively corresponds to $\sigma_{\rm ta}$ and $\sigma_{\rm c}$ from the $\Lambda$CDM cosmology. The colour of a shaded region at a fixed $z$ matches with the line colour at that $z$. The vertical dotted line is the resolution limit ($R=280$ kpc) of our simulation.}
    \label{fig:pk_vs_r}
\end{figure*}

\subsection{The smoothing scale (\texorpdfstring{$R$}{R}) dependence of the MPS}
\label{result:ps_m_smr}
$\Delta^2_{m}$ as a function of $R$ and for $k = 0.16 \mp$, which represent a sufficiently large scale, is shown in Figure \ref{fig:pk_vs_r} for halos (top panels) and \HI\ (bottom panels) fields. The left and right panels respectively are for $p_{\rm hd}$ and $p_{\rm ld1}$. For the clarity of presentation, we consider here only three different redshifts $z=1$, $2$, and $3$. Each line in a panel is plotted for $30$ different values of $R$ in the range $0$ to $4.2$ Mpc, at the step of $0.14$ Mpc. At first glance, we see that the MPS shows a strong dependence on the smoothing scale $R$. On the other hand, the standard \HI\ power spectra are expected to be independent of the smoothing scale 
as long as the smoothing scale is $\lesssim 2\pi/k_f$ where $k_f$ is the $k$ value where the power spectrum is considered, which is $0.16 \mp$ here as mentioned above.

\subsubsection{MPS for \texorpdfstring{$p_{\rm hd}$}{p-2}}

Considering the left panels, we see that with $p_{\rm hd}$, the shape of $\Delta^2_{m}$ for \HI\ field closely resembles that of the halo field, which is expected, as the \HI\ distribution traces halo distribution. However, the amplitude of $\Delta^2_{m}$ for \HI\ is smaller than that of the halo field as several peaks in $m$ which are present in the halo distribution are missed in the \HI\ distribution due to the \HI\ assignment prescription in eq.~\ref{eq:mh_mhi}. We further find that at a fixed $z$, the amplitude of $\Delta^2_{m}$ falls exponentially with increasing $R$ for both halo and \HI\ fields. This is also what we expect as with the increasing $R$, the peaks in their marked distribution get smoothed out, which results in the reduced amplitude in $\Delta^2_{m}$. Furthermore, $\Delta^2_{m}$ vs $R$ lines at $\Delta z=1$ are nearly parallel, but separated by a constant factor. Also, this separation increases with decreasing $z$. This difference for the \HI\ field is smaller compared to that for the halo field. This also matches our expectation as the evolution in the \HI\ overdensity is weaker than the halo overdensity.

\subsubsection{MPS for \texorpdfstring{$p_{\rm ld1}$}{p1}}
In the right panels of Figure \ref{fig:pk_vs_r}, $\Delta^2_{m}$ for $p_{\rm ld1}$, probes the clustering in the halo and \HI\ fields due to the distribution and evolution of the matter in the low-density regions. The trend of $\Delta^2_{m}$ for the \HI\ fields matches that for the halo field. However, the features are more clearly visible in the MPS of the \HI\ field than in the halo field for lower values of $R$. We now focus on $\Delta^2_{m}$ for the \HI\ field. We find that for small values of $R$, the amplitude of $\Delta^2_{m}$ decreases with decreasing $z$ at a fixed $R$. This trend gets reversed for large $R$ values, resulting in $\Delta^2_{m}$ vs $R$ lines for different redshifts intersecting each other at a point where $R \sim 0.8$ Mpc. As we have already established in section \ref{result:ps_m1} for $R=280$ kpc that $\Delta^2_{m}$ captures the low $m$ (or overdensity) regions along the web structures where $m$ decreases as the matter flows in with time (see the third row of Figure \ref{fig:m1_MW16}). This is indicated by the decreasing amplitude of $\Delta^2_{m}$ with decreasing $z$. Therefore, for $280\, {\rm kpc} \lesssim R \lesssim 0.8$ Mpc, $\Delta^2_{m}$ captures the evolving high-density regions with time along the web structures due to the flow of matter. We have also established in section \ref{result:ps_m1} that for $R=1.54$ Mpc, the amplitude of $\Delta^2_{m}$ increases with decreasing $z$, and this captures the regions with high $m$ (or lower-density) growing in size and number with time due to the flow of matter away from these regions. This flow makes nearly zero change in low $m$ ($<1$) regions for this $R$ value. Hence, the reverse trend, seen at $R>0.8$ Mpc where the amplitude of $\Delta^2_{m}$ grows with decreasing redshift, is largely due to the growth of low-density regions.

Further, we find that for a fixed $z$ (say $z=1$), there is a turnover value of $R\sim1.2$ Mpc, both sides of which the $\Delta^2_{m}$ values fall. For $R$ values smaller than this scale, the growth of $\Delta^2_{m}$ is rapid. On the other hand, for $R > 1.2$ Mpc, the decline of $\Delta^2_{m}$ with $R$ is relatively shallower. As we previously discussed, the growth of $\Delta^2_{m}$ with $R$ in the range $280\, {\rm kpc} \lesssim R \lesssim 0.8$ Mpc is decided by the evolution of the high-density regions along the cosmic web, this indicates that $\Delta^2_{m}$ continues to capture this evolution till $R \sim 1.2$ Mpc. The peak $R$ value is smaller for higher redshifts, and the distance between two successive peaks decreases with increasing $z$. Also, the peak $\Delta^2_{m}$ value decreases with increasing $z$. This indicates $\Delta^2_{m}$ to capture the growth of structures along the cosmic web with a faster rate at lower $z$ than that at higher $z$ and this growth is visible for $R$ values below the turnover scale. However, for $R$ values greater than the turnover scale, the evolution of the cosmic web is not visible as any change within the smoothing scale is being smoothed out, and the evolution of $\Delta^2_{m}$ here is controlled by the low-density regions. The faster decline of $\Delta^2_{m}$ with increasing $R$ indicates that the average size of the low-density regions is smaller at high redshifts, therefore, any changes within those at high $z$ can be masked easily by choosing higher $R$. At low redshifts, the low-density regions are relatively larger and therefore need larger $R$ to mask out the evolution.

The above results are found to be true for other $k$ values in the range $0.05\leq k \leq 1.0$ Mpc$^{-1}$. However, the nature of $\Delta^2_{m}$ vs $R$ curves change when a different value for $\rho_{\ast}$ is assumed or a different negative value for $p$ is taken. However, within a smaller range $0 < \rho_{\ast} \leq 0.1$, we find that the above results qualitatively hold.

We next investigate if the features of $\Delta^2_{m}$ vs $R$ for $p_{\rm ld1}$ in Figure \ref{fig:pk_vs_r} which can give more information about the growth of structures in the \HI\ field. The linear perturbation theory predicts that the growth of the structures in the self-gravity falls in the non-linear regime when the density fluctuation $\sigma (R,z)$ inside a sphere of radius $R$ attains a value between the turnaround and collapse values i.e., $\sigma_{\rm ta}=1.06$ and $\sigma_{\rm c}=1.686$ from the $\Lambda$CDM cosmology. The $\sigma_{\rm ta}$ is the fluctuation level at which the non-linear regime begins and the smoothing scale $R$ for which the average fluctuation of the field attains this value can be designated as $R_{\rm ta}$. On the other hand, at $\sigma_{\rm c}$ the matter for smoothing scale $R_{\rm c}$ gets collapsed to form bound objects. At each $z$, we plot a shaded region between two different $R$ values that correspond to the $\sigma_{\rm ta}$ (right edge) and $\sigma_{\rm c}$ (left edge) in the right panels of Figure \ref{fig:pk_vs_r}. The width of the shaded regions increases with decreasing $z$ and also the shaded region shifts toward higher $R$ with decreasing $z$. Considering the right-lower panel, at a fixed $z$, the part of $\Delta^2_{m}$ vs $R$ line falling in the shaded region probes the non-linear evolution of the \HI\ over the $R$-range corresponding to the shaded region. Also, the left edge of a shaded region of a particular colour intersects $\Delta^2_{m}$ vs $R$ line of the same colour at a point. This point moves in the direction of increasing amplitude of $\Delta^2_{m}$ with increasing $R$ and with decreasing $z$ which indicates capturing the hierarchal growth of the structures, where the small-scale structures are expected to form at high redshifts while at low redshifts larger size of structures is also expected.

\section{Summary and discussions}
\label{sec:summary}
The intensity mapping of the neutral hydrogen (\HI) distribution in the post-reionization era presents a potential means to investigate the \HI\ clustering and its evolution in this era. A number of models of \HI\ distribution in the post-reionization era indicate that \HI\ in this era is mostly associated with the galaxies that reside inside the dark matter halos which are sitting at the matter over-densities. That is why the 21-cm signal from this era is expected to be a cleaner probe of underlying matter clustering. However, the normal \HI\ density power spectrum only captures the information inherent in the high densities. The information about the \HI\ distributed at low and intermediate densities is missed in the standard \HI\ power spectrum. Many previous studies have shown that there is almost no evolution of the \HI\ clustering with time, indicated by the overlapping feature of its power spectrum at different redshifts over a wide range of $k$-modes. It is possible to perform a simple modification in the \HI\ density field in a physically meaningful and informative way so that the power spectrum from the modified field could capture the information about \HI\ distribution at different densities as well as its evolution with time. This modification serves as the mark and the modified field is denoted as the marked \HI\ field. Therefore, the power spectrum extracted from the marked field is known as the marked power spectrum (MPS). This work is the first application of the MPS on the \HI\ distribution in the post-reionization era and focuses on studying the \HI\ clustering and its evolution at various length scales and redshifts in this era.

In this work, we rely on a semi-numerical technique to simulate the \HI\ density field in the redshift range $1 \leq z \leq 5$. We then apply a non-linear transformation (eq.~\ref{eq:m_mw16}), formulated in ref. \citep{white16}, on the fields to produce the marked field $m$. This non-linear transformation is a function of local density only and has three free parameters. These parameters are the smoothing radius $R$, a threshold on the local density $\rho_{\ast}$, and an exponent $p$. The values of these parameters are chosen based on the features and length scales of interest. Therefore, by tuning the various parameters, we can extract information from various regions using MPS, as opposed to the regular \HI\ density power spectrum, which mostly carries information from the high-density regions. Further, since $m$ is a non-linear function of density, MPS contains some higher-order density statistics and hence retains more information compared to the regular density power spectrum. Here, we have studied MPS for a number of different parameter combinations and compared the results with the regular density power spectrum. We have also produced different visual maps of normal and marked \HI\ density fields and tried to interpret the results exhibited by the power spectrum. The main findings from our analysis are:

\begin{itemize}
    \item Our simple semi-numerical setup suggests that there is an evolution in \HI\ density at low and intermediate-density regions ($\rho/\bar{\rho}<3$) that are not captured by the standard density power spectrum.

    \item Considering marks $m \propto [1/\rho_{\rm  R}]^p$, we find that for a positive $p$, MPS can capture the evolution of the low-density regions, whereas for a negative $p$ the mark is sensitive to the evolution of the high-density regions similar to the standard density power spectrum. However, $m$ with $p<-1$ tends to enhance even the slightest evolution in the high-density regions. As a result, MPS with $p<-1$ shows evolution with $z$, unlike the standard power spectrum. Although the results are sensitive to the choice of the smoothing radius $R$.

    \item MPS of the \HI\ field shows an approximate power law behaviour with $k$-mode at every $z$ similar to the normal \HI\ power spectrum, although having a different slope. This indicates that mark fields can also be viewed as biased tracers of the underlying matter field.

    \item Considering our exact mark function in eq.~\ref{eq:m_mw16} and for a fixed $\rho_{\ast}=0.01$, we find that for $p=-2$, the MPS shows significant redshift evolution at all $k$ for the smallest value of $R$. As $R$ increases, the $z$ evolution tends to become less prominent, and finally, for the highest value of $R$ we find that MPS evolves only marginally with $z$. This is because, with the increasing $R$, we are smoothing out the density field. When $R$ becomes comparable to the typical size of the over-dense regions, any changes happening within those will be masked out. As a result, the information about the evolution will not be fully captured by MPS.

    \item Now, for $p=1$, we find that the $z$ evolution of MPS is significant across the various $R$ values chosen. Only the order of the evolution with $z$ changes between small and large $R$, causing a cross-over point at $R\sim0.8$ Mpc where the results for different $z$ intersect and we find a minimum evolution of MPS around this scale. For a fixed $z$ (say $z=1$), we find a turnover value around $R\sim1.2$ Mpc. To the left of this point, MPS increases rapidly as $R$ increases capturing the evolution of the high-density regions in the cosmic web due to the transfer of matter from low to high densities. To the right of this point, the MPS gradually declines with increasing $R$ capturing the evolution in the low-density regions only. The typical size of the low-density regions is larger than the size of the high-density regions. As a result, the evolution in the low densities can still be captured even at the highest value of $R$ chosen in the analysis. The size of these regions grows with decreasing $z$. Therefore, we expect to capture the evolution of the low-density regions at low $z$ using MPS even if the field is heavily smoothed.
    
    \item We conclude that the evolution of \HI\ density, driven by the flow from low- to high-density regions, can be effectively probed across different density regimes and stages by appropriately selecting the functional form and parameters of the mark. MPS may also reflect information about some higher-order statistical moments of the \HI\ field, thereby offering a richer set of statistical insights. Finally, information about the \HI\ content at low and intermediate densities is crucial for a correct and consistent analysis of \HI\ content and its evolution, particularly when using the 21-cm background. MPS offers a less biased statistic for constraining parameters compared to the standard power spectrum.
       
\end{itemize}

This work is the first to consider the application of mark statistics on the post-reionization \HI\ field to track the distribution and evolution of the \HI\ content using the corresponding marked power spectra. The current analysis is preliminary as we consider only a very simple model of \HI\ distribution where \HI\ resides at the center of each dark matter halo and we focused mostly on one functional form of the mark, albeit with various values of the parameters. In practice, we expect the \HI\ to be contained by galaxies that reside inside the halos. Some diffuse \HI\ may reside outside of the galaxies as well. Not only that but \HI\ also moves as a combination of bulk and local motions \cite{Villaescusa-Navarro:2018vsg}. These collectively contribute to shaping the \HI\ distribution in the post-reionization era. Furthermore, our simulation cannot resolve the \HI\ containing halos below a certain mass of $1.09\times10^9\, M_{\odot}$ at $z>3.5$. The \HI\ densities corresponding to these halos will likely contribute to low- and intermediate-density regimes. However, ref.~\citep{sarkar16} discussed that ignoring the smaller halos has a minor effect on the \HI\ power spectrum even at $z \sim 6$ but may have a significant impact on MPS. We plan to use more robust simulations and address all the above issues in future work.

%In order to quantify the contributions from the higher correlations and their information content, eq. \ref{eq:m_mw16} does require the Taylor expansion followed by quantifying the contributions from each term in the expansion for a set of fixed mark parameters, which is beyond the scope of this paper.

\section{Acknowledgements}
MK and MS acknowledge financial support from the foundations Carl Tryggers stiftelse för vetenskaplig forskning (grant agreement no. CTS 21:1376) and Magnus Bergvalls stiftelse (grant agreement no. 2021-04407) awarded to docent Martin Sahlén. MS also acknowledges financial support from the Swedish National Space Agency (Rymdstyrelsen) through the Senior Researcher Career grant No. 2020-00108. The simulations and statistical analysis presented in this paper have mainly used the computing resources available to the Cosmology with Statistical Inference (CSI) research group at the Indian Institute of Technology Indore (IIT Indore). The computations and data storage were also enabled by resources provided by the National Academic Infrastructure for Supercomputing in Sweden (NAISS), partially funded by the Swedish Research Council through grant agreement no. 2022-06725. DS acknowledges the support of the Canada $150$ Chairs program, the Fonds de recherche du Qu\'{e}bec Nature et Technologies (FRQNT) and Natural Sciences and Engineering Research Council of Canada (NSERC) joint NOVA grant, and the Trottier Space Institute Postdoctoral Fellowship program. SM acknowledges financial support through the project titled “Observing the Cosmic Dawn in Multicolour using Next Generation Telescopes” funded by the Science and Engineering Research Board (SERB), Department of Science and Technology, Government of India through the Core Research Grant No. CRG/2021/004025.

%%%%%%%%%%%%%%%%%%%%%%%%%%%%%%%%%%%%%%%%%%%%%%%%%%

%%%%%%%%%%%%%%%%% APPENDICES %%%%%%%%%%%%%%%%%%%%%

\appendix

\section{Appendix}
\label{app:app1}

\subsection{A different functional form of the mark}
\label{app:mark_W09}
We have already discussed in section \ref{sec:formalism} that the mark could be any arbitrary function of the property of a field. However, it is chosen according to the information in the field that we want to draw and which is not straightforward to extract from the standard distribution. Since, in this paper, we are interested in studying the \HI\ clustering and its evolution with time. This is why we opted for the mark as a function of local density which is smoothed over a scale $R$ and used the formulation of ref. \citep{white16} (eq. \ref{eq:m_mw16}). The major advantage of using eq. \ref{eq:m_mw16} is that this equation for $\rho_{\ast} \rightarrow 0$ scales as $m \approx [1/\rho_R]^p$ and when it is applied to a field, then the power spectrum of this field does explicitly feature the density. Another functional form of the mark as a function of local density is formulated in ref. \citep{white09} and is given by
\begin{equation}
   m_1(\textbf{x},t; \rho_{\ast}, p) = \frac{[\rho(\textbf{x},t)]^p}{[\rho_{\ast}]^p+[\rho(\textbf{x},t)]^p}\, ,
    \label{eq:m_mw09}
\end{equation}
This form was used to encode the information about which galaxies reside inside which halos.
This is because the $m_1$ produces the marked field in the range $0$ (for $\rho \ll \rho_{\ast}$) and $1$ (for $\rho \gg \rho_{\ast}$) in the sense that a given position either has a galaxy or not. Hence, the marked power spectrum with $m_1$ does not explicitly feature a density.  Note that the $m_1$ is the function of non-smoothed density, whereas the observable of the \HI\ field is expected to be related to the smoothed \HI\ density, which is a continuous variable and not $1$ or $0$. Thus, the $m_1$ is not informative in the sense of probing the \HI\ clustering and its evolution with time in the post-reionization era.

\subsection{The impact of the parameter \texorpdfstring{$\rho_{\ast}$}{rho} on the marked power spectra}
\label{app:ps_m1_diif_rho}

\begin{figure*}
    \centering
    \includegraphics[width=1\textwidth,angle=0]{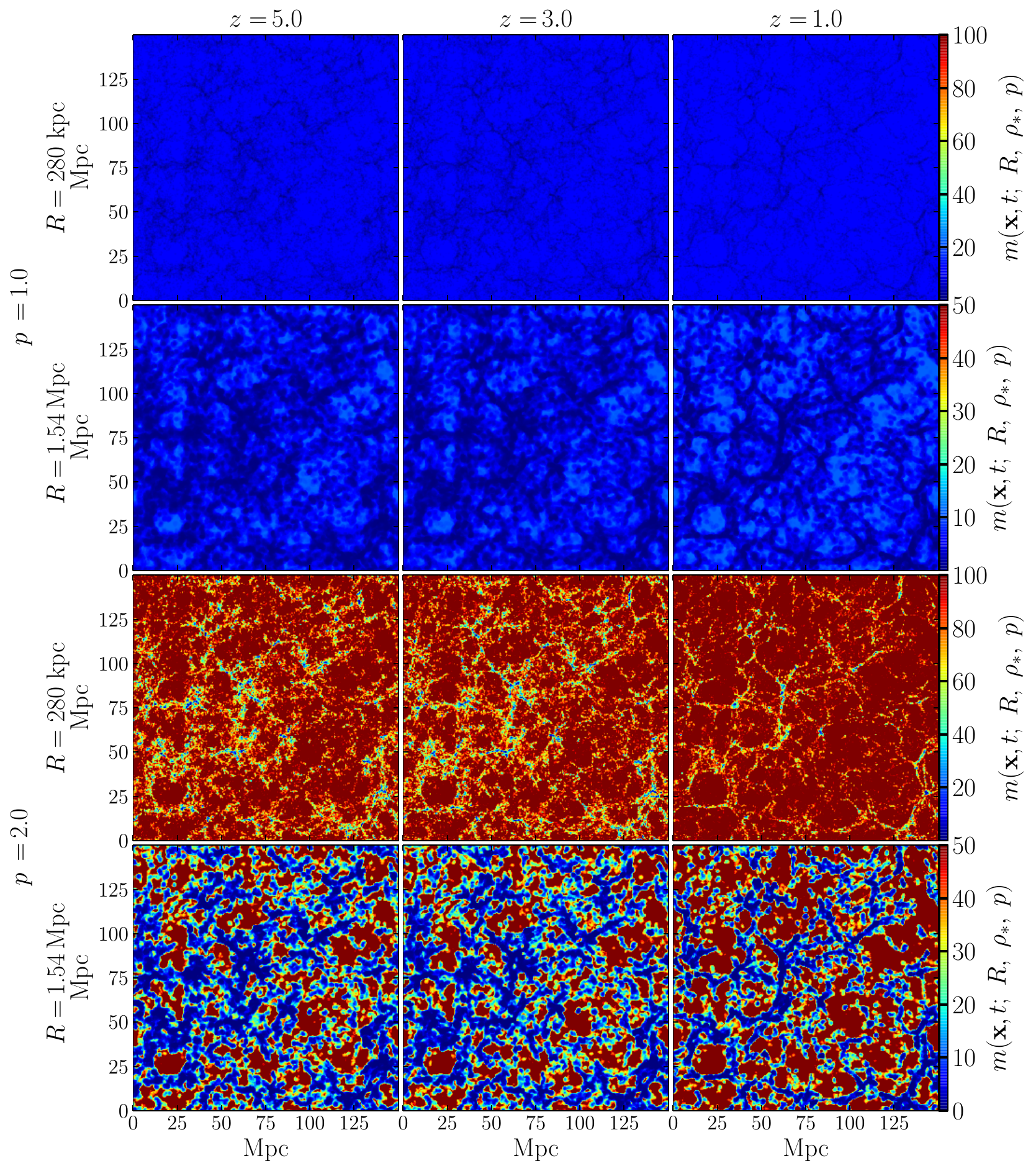}
    \caption{Marked \HI\ field ($m$) at different redshifts (decreasing from left to right) and for different mark parameters (varying in the vertical direction). Here, the value of the parameter $\rho_{\ast}$ is $0.1$. The top two rows show the time evolution in $m$, calculated for two different $R$ values i.e., $R=280$ kpc and $1.54$ Mpc as well as $p_{\rm ld1}$. The bottom two rows show the same but now for $p_{\rm ld2}$.}
    \label{fig:m1_MW16_diff_roast}
\end{figure*}

\begin{figure*}
    \centering
    \includegraphics[width=1\textwidth,angle=0]{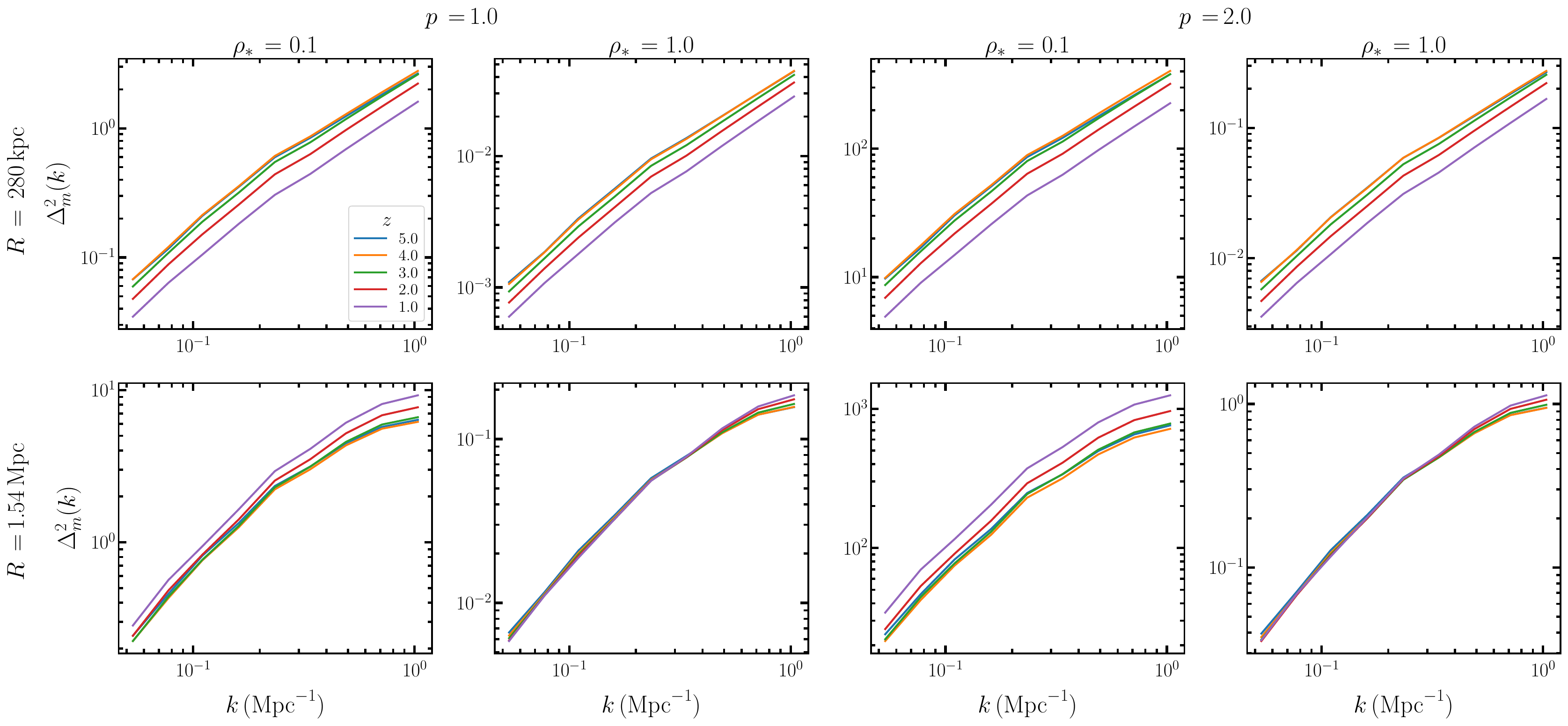}
    \caption{Marked \HI\ power spectra ($\Delta^2_{m}$) as a function of $k$ mode at five different redshifts. The upper panels are shown at a smoothing scale of $R=280$ kpc, whereas the lower panels are shown at $R=1.54$ Mpc. The two left-columns are at $p_{\rm ld1}$ and for two different $\rho_{\ast}$ i.e., $\rho_{\ast}=0.1$ and $1.0$ whereas the two right-columns are at $p_{\rm ld2}$.}
    \label{fig:pk_MW16_diff_roast}
\end{figure*}

The figures in this paper are shown for a fixed value of the free parameter $\rho_{\ast}$ i.e., $\rho_{\ast}=0.01$. It is, however, imperative to look at the impact of different values of this parameter on the marked power spectrum in order to see the robustness of the results. In Figure \ref{fig:pk_MW16_diff_roast}, we show the marked power spectrum ($\Delta^2_{m}$) at various redshifts for different values of the marked parameters $\rho_{\ast}$, $p$ and $R$. We first consider the two left columns at $\rho_{\ast}=0.1$ and $1.0$ and with $p_{\rm ld1}$. Comparing these two columns with the middle column of Figure \ref{fig:pk_MW16}, which shows $\Delta^2_{m}$ for $\rho_{\ast}=0.01$ and $p_{\rm ld1}$, side by side, we see that for each smoothing scale the amplitude in $\Delta^2_{m}$ decreases with increasing $\rho_{\ast}$. The shape of $\Delta^2_{m}$, however, remains the same. Here also, for a fixed $\Delta z$, the separation between two $\Delta^2_{m}$ vs $k$ lines decreases with increasing $\rho_{\ast}$. In order to understand these behaviours, we compare the marked field in the two bottom rows of Figure \ref{fig:m1_MW16}, which show the marked field for $\rho_{\ast}=0.01$ and $p_{\rm ld1}$, with the two top rows in Figure \ref{fig:m1_MW16_diff_roast} at $\rho_{\ast}=0.1$ and $p_{\rm ld1}$. We find that with a larger value of $\rho_{\ast}$ the marked field shows its lower amplitude as well as the decreased contrast between high and low $m$. Due to this, $\Delta^2_{m}$ amplitude decreases with increasing $\rho_{\ast}$. Also, the evolution in $m$ can not be clearly visualized in the marked field with larger $\rho_{\ast}$. This evolution is also not captured well statistically by $\Delta^2_{m}$.

Now, considering the two right columns in Figure \ref{fig:pk_MW16_diff_roast}, we find that the larger value of $p$ (here $p=2$) increases the amplitude of $\Delta^2_{m}$. The larger $p$ also increases the separation between two $\Delta^2_{m}$ vs $k$ lines for a fixed $\Delta z$ capturing the redshift evolution in the marked field. However, the increase in this separation is substantial for $\rho_{\ast}=0.1$. This can be visualized in the two bottom rows in Figure \ref{fig:m1_MW16_diff_roast} at $\rho_{\ast}=0.1$ and $p_{\rm ld2}$. Comparing these rows with two top rows, we see the increased amplitude in $m$ for $p_{\rm ld2}$ as well as the increased contrast between high and low $m$. Here also, the evolution in $m$ is visually clear which is well captured by $\Delta^2_{m}$. We thus conclude that with a small $p$, the small value of $\rho_{\ast}$ is a good choice to probe the \HI\ clustering and its evolution with time.

%The analysis for different choices of the $\rho_{\ast}$ has been made to investigate also the relevant values of this parameter which makes the mark quite sensitive for the study of the post-reionization era using the \HI\ $21$-cm signal.

%\subsection{Vizualization of \HI\ distribution in the standard and marked distribution of different resolutions}
%\label{app:h1_map_diff_r}

%\begin{figure*}
    %\centering
    %\includegraphics[width=1\textwidth,angle=0]{Plots/m_maps_at_different_radius.pdf}
    %\caption{2D slices of the \HI\ distribution in the standard (top panels), and marked (middle panels) density fields for four different smoothing radii and at redshift $z=1$. The bottom panels show how much the marked \HI\ distribution is temporally evolved by $z=1$ relative to the distribution at $z=3$ in the different smoothed environment. For this we plot the quantity $\Delta m({\bf x},z)=m({\bf x},z=1)-m({\bf x},z=3)$ at different $R$. The other mark parameter values are fixed and have to be taken $p=1.0$ and $\rho_{\ast}=0.01$.}
    %\label{fig:maps_at_diff_r}
%\end{figure*}

\bibliographystyle{JHEP}
\bibliography{reference}

% Don't change these lines
%\bsp	% typesetting comment
%\label{lastpage}
\end{document}